\documentclass[aps,prl,amsmath,amssymb,amsfonts,floatfix,nofootinbib,superscriptaddress,longbibliography,12pt,reprint,noeprint]{revtex4-2}

\usepackage{geometry}
\usepackage[english]{babel}
\usepackage{blindtext}
\usepackage{amsmath,amsfonts,amssymb}

\usepackage{graphicx}
\usepackage{float}
\usepackage{times}
\usepackage{color}

\usepackage{subfigure}
\usepackage{indentfirst}
\usepackage{setspace}
\usepackage{bm}

\usepackage[colorlinks,linkcolor=blue,anchorcolor=blue,citecolor=green]{hyperref}
\usepackage{CJKutf8}
\newcommand{\Tr}{\mathrm{Tr}}

\graphicspath{{fig}}

\allowdisplaybreaks
\newcounter{amgg}
\newcounter{fan}
\newcounter{yc}

\begin{document}
\title{No measurement induced phase transition in the entanglement dynamics of monitored non-interacting one-dimensional fermions in a disordered or quasiperiodic potential}
\author{Can Yin (\begin{CJK*}{UTF8}{gbsn}殷灿\end{CJK*})$^*$}
\email{yin\_can@sjtu.edu.cn}
\affiliation{Shanghai Center for Complex Physics, School of Physics and Astronomy,
\\Shanghai Jiao Tong University, Shanghai 200240, China}
\affiliation{Tsung-Dao Lee Institute, Shanghai Jiao Tong University, Shanghai 201210, China}
\author{Bo Fan (\begin{CJK*}{UTF8}{gbsn}范波\end{CJK*})$^*$}
\email{bo.fan@sjtu.edu.cn}
\affiliation{\mbox{Institute for Quantum Materials and Technologies, Karlsruhe Institute of Technology, 76131 Karlsruhe, Germany}}
\affiliation{\mbox{Institut f\"ur Theorie der Kondensierten Materie, Karlsruhe Institute of Technology, 76131 Karlsruhe, Germany}}
\affiliation{Shanghai Center for Complex Physics, School of Physics and Astronomy,
	\\Shanghai Jiao Tong University, Shanghai 200240, China}
\author{Antonio M. Garc\'ia-Garc\'ia}
\email{amgg@sjtu.edu.cn}
\affiliation{Shanghai Center for Complex Physics, School of Physics and Astronomy,
\\Shanghai Jiao Tong University, Shanghai 200240, China}
\vspace{0.2cm}
\date{\today}
\vspace{0.3cm}

\begin{abstract}
	We show that the entanglement entropy (EE) of one-dimensional (1d) non-interacting fermions with $U(1)$ symmetry in the presence of a disordered or quasi-periodic potential in which the occupation number is being monitored by homodyne or projection protocols is always in an area-law phase so no measurement induced phase transition (MIPT) occurs. The reason for the previously claimed MIPT in these systems was a finite size effect related to the fact that the maximum lattice size $L \simeq 500$ was of the order of the correlation length. By increasing the system size up to $L \simeq 18000$, employing Graphics Processing Units (GPUs), and performing a careful finite-size scaling analysis, we find that the critical monitoring strength is consistent with zero, so no MIPT occurs. For the disordered case, these numerical results are fully supported by an analytical calculation based on mapping the problem onto a nonlinear sigma model (NLSM) that confirms the absence of the MIPT for any monitoring or disorder strength. The effect of disorder is captured by a change of symmetry from BDI to AIII, which results in an enhanced correlation length in the weak disorder limit, and by an effective monitoring strength that increases linearly with disorder.
\end{abstract}

\maketitle
\def\thefootnote{*}\footnotetext{These authors contributed equally to this work}
Disorder has a profound impact on the dynamics of quantum mechanical systems. Quantum coherence effects in one (1d) and two dimensions (2d) induce  Anderson localization \cite{anderson58,abrahams1979} for any strength of the disordered potential. In this insulating phase, eigenstates are exponentially localized and spectral correlations are described by Poisson statistics. 
In three and higher dimensions an Anderson metal-insulator transition \cite{schreiber1991,rodriguez2012,garcia2007dimensional,rodriguez2011multifractal,garcia2008a} occurs at a finite disorder strength. The metallic phase is characterized by states delocalized in space and level statistics described by random matrix theory. 
Similarly, a metal-insulator transition takes place \cite{harper1955,aubry1980,kohmoto1983} in a 1d system in the presence of a quasiperiodic potential. Disorder has also a strong impact on the entanglement properties. For instance, the entanglement entropy (EE) of the ground state of 1d non-interacting fermions obeys an area law if disorder is present while it scales logarithmically \cite{Burmistrov2017,gloev2006} with the system size in the clean limit. 

In recent years, motivated by both \cite{warren1986,zoller1987,gleyzes2007,minev2019,plenio1998,wiseman1993,collett1987,wiseman2014,fuwa2015} experimental advances in quantum optics and its relevance in quantum computation, there has been a growing interest in the dynamics of monitored quantum systems, namely, systems that are being measured. Although research was initially focused on single particle and systems with few levels, there is growing interest in the effect of monitoring in the more realistic case of many-body quantum systems which brings new challenges and opportunities. One of the most interesting and more intensively investigated problems is that of the entanglement dynamics and, more specifically, the conditions for the existence of measurement induced phase transitions \cite{Li2018a,Skinner2019a,nahum2021a,ippoliti2021a,Li2019a,carisch2023,Szyniszewski2019a,Szyniszewski2020,Turkeshi2020a,Turkeshi2021,Turkeshi2022a,legal2023,Soares2024} (MIPT) where for sufficiently long times the scaling of the EE, or a similar observable, with the system size undergoes a qualitative change, say from volume to area, by tuning the monitoring strength. The existence of this type of transition has been confirmed experimentally in a few cases \cite{Noel2022a,Koh2022} though we are far from a full experimental control. 
In a many-body context, the numerical simulation, or the analytical description, of the entanglement dynamics of monitored systems is in general challenging. Phenomenological random unitary circuits \cite{Li2018a,Li2021,Jian2020a,Jian2022,Jian2023} and non-interacting fermions,  \cite{poboiko2023,poboiko2023a,buchhold2021a,buchhold2022,fava2023,fava2024,cao2019a, poboiko2026, poboiko2025a,chaki2026} are two exceptions where explicit results can be obtained for large sizes both analytically and numerically. For 1d Majorana fermions, with the parity being monitored, it was found \cite{fava2023} that a MIPT occurs at finite monitoring strength. By contrast, for 1d non-interacting fermions with a $U(1)$ conserved charge, the EE of the system \cite{poboiko2023a} is in the area law phase for any monitoring strength so no MIPT occurs. These findings were obtained by the mapping of the entanglement dynamics onto a non-linear sigma model (NLSM) \cite{wegner1979,wegner1987,efetov1983supersymmetry} which is solved by field theory techniques. For the latter, these results were recently confirmed by large scale numerical simulations involving Graphics Processing Units (GPU) \cite{garciagarcia2026}. A MIPT occurs for free fermions in higher spatial dimensions \cite{poboiko2023a,chahine2023}, by increasing sufficiently the hopping range \cite{minato2022,mueller2022,fuji2020}, by turning on interactions \cite{poboiko2024} or by employing non-commuting local measurements \cite{poboiko2025a}.

A natural question to ask is whether the presence of a disordered or quasiperiodic  potential, which, as mentioned earlier, dramatically changes the dynamics in the non-monitoring limit, has a strong effect on the entanglement dynamics of the monitored system. We note that quantum coherence effects leading to Anderson localization are quite fragile so they are likely destroyed or heavily suppressed in the presence of monitoring. 
This question was recently addressed in the case of 1d free fermions in a disordered \cite{lunt2022} and a quasiperiodic \cite{matsubara2025} potential. In both cases, it was found that a MIPT occurs at a finite strength of the potential. This is somehow surprising since, as mentioned earlier, no MIPT occurs in the clean case for any finite monitoring strength. One would expect that, as in the disordered no-monitoring case \cite{gloev2006,Burmistrov2017}, disorder reduces, not increases, the scaling of the EE with system size. 
In this paper, we show that indeed, the presence of a quasiperiodic potential or a random potential does not change the scaling of the EE with respect to the clean case. Therefore, for any finite monitoring or disorder strength, the scaling of the EE with system size, for sufficiently long times, is area law so there is no MIPT. The transition claimed in Refs.~\cite{lunt2022,matsubara2025} was a finite size effect related to the fact that the maximum lattice size was only $L \simeq 500$ while the correlation length could be much larger in the limit of weak monitoring and disorder strength. As a result, the observed deviations from the area law in this limit were due to the fact that the system size was smaller than the correlation length. In order to gather conclusive evidence of the absence of a MIPT in this setting, we will see it is necessary to reach $L \simeq 18000$ and perform a careful finite size scaling analysis. For the disordered case, the absence of a MIPT is demonstrated analytically by mapping the problem onto a NLSM and using field theory techniques. We find that due to a change in symmetry with respect to the clean case, the correlation length in the limit of weak disorder is larger and it increases with the disorder strength.  We begin by introducing the model and the monitoring protocols. 

{\it The model.-} We consider the one-dimensional free fermionic chain with on-diagonal potential $V_i$,
\begin{equation}
H = J \sum_{i=1}^{L} [ c_{i}^\dagger c_{i+1} + c_{i+1}^\dagger c_{i} ] +\sum_{i=1}^{L} V_i c_{i}^\dagger c_{i} ,
\label{eq:Dis_Ham}
\end{equation}
where $c_i$ and $c_i^\dagger$ are the annihilation and creation operators for fermions at site $i=1,2,\cdots L$ and we set $J=1$. The filling rate is $N/L=1/2$, with $N$ the number of fermions, and the initial state is $|\psi\rangle_0=|10101010\cdots\rangle$ with $0$, $1$ the site occupation numbers. We consider two potentials $V_i$:\par
(1) Disordered potential: $V_i$ is the random potential extracted from a box distribution between $[-W,W]$. We impose periodic boundary conditions $c_{i}=c_{i+L}, c_{i}^\dag=c_{i+L}^\dag$. \par
(2) Quasiperiodic potential: $V_i = V\cos({2\pi i \over \tau} + \theta)$, where $V$ is the strength of the quasiperiodic potential, and we consider the golden mean $\tau = (\sqrt{5}+1)/2$. We fix $\theta=0$ for simplicity. Since $\tau$ is an irrational value, we choose the system sizes $L$ from the Fibonacci numbers. This allows the quasiperiodic potential to be compatible with the periodic boundary conditions.

We consider the projective measurement (PM) protocol for the quasiperiodic potential and the homodyne measurement protocol, usually termed quantum state diffusion (QSD), for the disordered potential; see Appendix~\ref{app:protocol} for further details. These choices match the protocols employed in previous studies \cite{matsubara2025,lunt2022}.
For both protocols, the observable being measured is the occupation number $n_i =c_i^\dag c_i$. 
In the PM protocol, projective measurements of the occupation number of individual sites, chosen randomly, occur at random times whose frequency is governed by a Poisson distribution depending on a parameter $\gamma$. The outcome of each measurement projects the wavefunction onto an eigenstate of $n_i$ with eigenvalue either $0$ or $1$. The probabilities of the two outcomes are given by the Born rule as $\langle n_i\rangle$ and $1-\langle n_i\rangle$ for the filled and unfilled states at site $i$, respectively, and therefore depend on the state before the measurement event.
By contrast, in the QSD protocol, all sites are weakly measured at each time step. The state evolution is governed by a stochastic equation with a Gaussian noise of zero average and variance $\gamma dt$, where $dt$ is the time step.
We focus on the long-time steady-state behavior of the EE. It is possible  to expand the EE in a series of cumulants of the subsystem particle number. It was shown in Refs. ~\cite{levitov2009,poboiko2023} that the leading order in this expansion, the second cumulant determined by the connected density-density correlation function $C(r,t)$ is enough to describe the EE. Therefore, after the EE has saturated, we average $C(r,t)$ over the steady-state interval $t\in[L/2,L]$ and henceforth denote the resulting time-averaged correlation function by $C(r)$.
 
\begin{figure}[!htb]
	\centering
	\includegraphics[width=7.5cm]{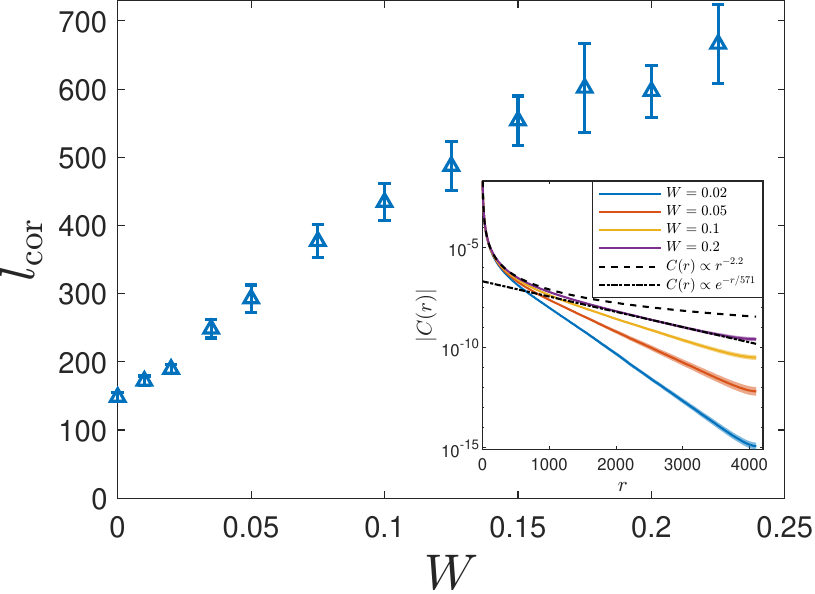}
	\caption{Correlation length $l_{\rm cor}$ as a function of the disorder strength $W \in [0.01,0.225]$ at a monitoring strength $\gamma=0.6$ for the QSD protocol. For $W=0$, we use the data in Fig. 2(b) of \cite{garciagarcia2026}. The data are averaged over at least three disorder realizations, at least ten quantum trajectories for each realization, and 33 equally spaced time points in the steady-state interval $t\in[L/2,L]$.
As shown, the correlation length $l_{\rm cor}$ increases with the disorder strength $W$, indicating that larger system sizes are necessary to identify the area-law phase. Inset: Absolute value $|C(r)|$ of the density correlation function for $W=0.02,0.05,0.1,0.2$, employing the same data set as in the main panel. The shaded band around each curve is the error bar (the 95\% confidence interval). For $W=0.2$, we perform two fits in different spatial regimes. The dashed line shows the power-law fit $C(r)\propto r^\alpha$ over $r\in[5,500]$, giving $\alpha\simeq -2.2$. The dash-dot line shows the exponential fit $C(r)\propto \exp(-r/l_{\mathrm{cor}})$ over $r\in[2500,3300]$, giving $l_{\mathrm{cor}}\simeq 570$.
}\label{fig:QSD_mu06}
\end{figure}

{\it Numerical Entanglement dynamics-}
For the QSD protocol corresponding to the disordered potential, we implement the bulk of our numerical operations on the GPU (device) using C++ and CUDA platforms, reserving the CPU (host) for only lightweight tasks that do not justify additional kernel launches. The PM protocol corresponding to the quasi-periodic potential relies on MATLAB's GPUArray framework to treat the GPU as an accelerator of the CPU workload, which is similar to CUDA when no fine-grained control over memory allocation or thread organization is needed.\\
We now proceed with the calculation of the correlation function $C(r)$ for the disordered potential using the QSD protocol. Our main goal is to provide sufficient numerical evidence to demonstrate the absence of an MIPT in a disordered quantum monitoring system. Later, this numerical evidence will be supported by an explicit analytical calculation. 
As mentioned earlier, it was shown in Refs.~\cite{poboiko2023,fava2024} that, for $V_i=0$, the entanglement dynamics after saturation can be mapped to a NLSM whose $SU(2R)/Sp(2R)$ target manifold is dictated by the BDI symmetry of the doubled Lagrangian operator.
The NLSM prediction for $C(r)$ \cite{poboiko2023,poboiko2024,fava2024} is, 
\begin{equation}\label{eq:Crlcor}
	C(r,t\to\infty)\sim
	\left\{
	\begin{aligned}
		&\exp(-r/l_{\mathrm{cor}}), && r\gg l_{\mathrm{cor}},\\
		&1/r^{2}, && r\ll l_{\mathrm{cor}}.
	\end{aligned}
	\right.
\end{equation} 
The asymptotic exponential decay implies that the EE obeys an area law for any $\gamma>0$, and we therefore focus on the correlation length $l_{\rm cor}$, which can be extracted from an exponential fit to $C(r)$ only when $L\gg l_{\rm cor}$, as illustrated in the inset of Fig.~\ref{fig:QSD_mu06}.

Importantly, the correlation length \cite{poboiko2023,poboiko2024,fava2024} in the clean case, 
\begin{equation}
	l_{\rm cor}\sim 
		\frac{1}{\gamma}\exp\left(\frac{\sqrt{2}\pi}{2\gamma} \right), \text{BDI, No disorder},
	\label{eq:RGBDI}
\end{equation} 
that belongs to the BDI \cite{poboiko2023,poboiko2024,fava2024} universality class, increases exponentially with $1/\gamma$. Therefore, for $\gamma \ll 1$, it is challenging to confirm numerically the area-law phase because for weak monitoring $l_{\rm cor} \gg L$, so only the power-law decay, related to a critical-law phase, is observed. 
However, the latter is a finite size artifact. A similar problem occurs in the 2d Anderson model \cite{mackinnon1983,eilmes1998} where it is necessary to reach $L^2\sim 100^2$ to show the analogue result that, for any disorder, all eigenstates are exponentially localized. It has been recently shown \cite{garciagarcia2026} that system sizes of order $L\sim 10^4$ are required to determine the existence, or not, of a MIPT by extrapolating the numerical small $\gamma$ results to $\gamma \to 0$. We will follow this route here. 

In the presence of disorder, we shall see that the symmetry is AIII rather than BDI~\cite{Mantikupblished}. We consider a small disorder strength $W$, which is sufficient to break the symmetry. For $\gamma\gg w$ and a nonzero disordered potential $V_i\neq0$, we expect,
\begin{equation}
l_{\rm cor}\sim
\frac{1}{\gamma}\exp\left(\frac{\sqrt{2}\pi}{\gamma}\right), \text{AIII},
\label{eq:RGAIII}
\end{equation}
which differs by a factor of two in the correlation-length exponent from the clean-limit result in Eq.~(\ref{eq:RGBDI}). This result follows from mapping the entanglement dynamics onto an $SU(R)$ NLSM, which also predicts a shift $\gamma\to\gamma+w$ that becomes non-negligible only when $\gamma/w\sim O(1)$. Note that for the analytical calculation we shall use Gaussian disorder of zero mean and variance $w=W^2/3$.

In Fig.~\ref{fig:QSD_mu06}, we show $l_{\rm cor}$ as a function of the disorder strength $W$ at fixed monitoring strength $\gamma=0.6$ and $L=8192$. We perform averaging first over quantum trajectories, then over disorder realizations. The ensemble sizes are chosen according to the size of the error bars, which are defined from the $95\%$ confidence interval, resulting in about $36$-$52$ trajectories in total for each point. We note that, in the legend of the inset of Fig.~\ref{fig:QSD_mu06}, the correlation length, defined from the exponential decay rate of $C(r)$, is computed by first averaging $C(r)$ over quantum trajectories and then fitting the resulting function, whereas in the rest of the paper we first compute $l_{\mathrm{cor}}$ for each trajectory and then average the results. We observe that the correlation length increases with disorder, making the exponential decay of $C(r)$, which signals the area-law phase, more difficult to identify than in the clean limit. Even in the clean limit, system sizes of order $L\simeq 10000$ were required to extrapolate the numerical results to $\gamma\to0$ \cite{garciagarcia2026}.

In order to determine whether a MIPT occurs, we fit the numerical values of $l_{\rm cor}$ to the form,  
\begin{equation}
	l_{\rm cor}= \frac{b}{|\gamma-\gamma_c|}\exp\left(\frac{a }{|\gamma-\gamma_c|} \right),
	\label{eq:lcorfit}
\end{equation} 
where the exponential coefficient $a$ and the prefactor $b$ are treated as fitting parameters. We also introduce a constant shift $\gamma_c$ in $\gamma$ as an additional fitting parameter. This fitting function assumes that in the weak-disorder regime $\gamma\gg w$ the shift $\gamma\to\gamma+w$ can be neglected. We will show later that this is the case. 
To account for the uncertainties in $l_{\rm cor}$ in the fitting procedure, we use weighted least squares, which minimizes the residuals weighted by the inverse variance.

\begin{figure}[!htb]
	\centering
	\includegraphics[width=7.5cm]{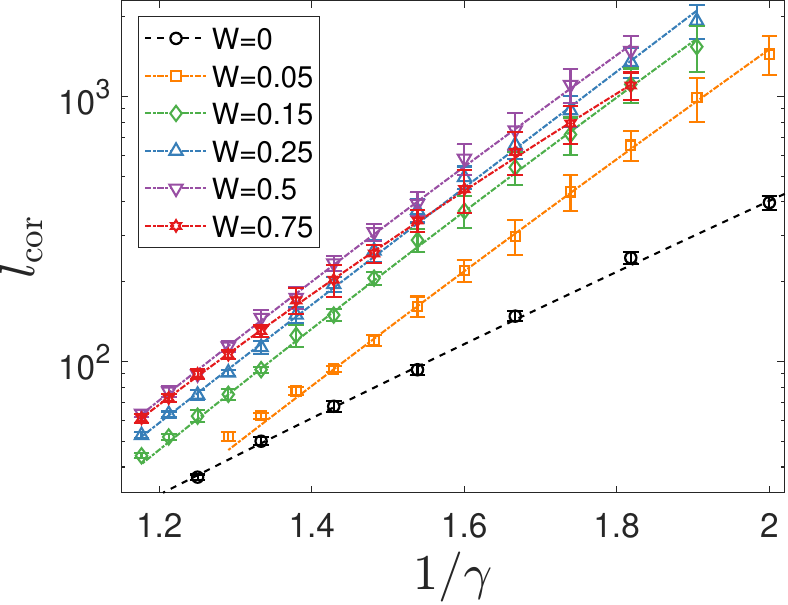}
	\caption{Correlation length $l_{\rm cor}$ as a function of the measurement strength $\gamma$ for the disordered system under the QSD protocol. For most data points, we use $L=8192$; when the correlation length approaches the system size, we instead use $L=10000$ for $(W,\gamma)=(0.15,0.55),(0.25,0.55),(0.5,0.575),(0.75,0.55)$, $L=14000$ for $(W,\gamma)=(0.15,0.525),(0.5,0.55)$, and $L=18000$ for $(W,\gamma)=(0.25,0.525)$. The dash-dotted lines are fits to Eq.~(\ref{eq:lcorfit}) over $\gamma\leq0.7$ for $W=0.05$, $\gamma\leq0.8$ for $W=0.15$, and $\gamma\leq0.85$ for $W=0.25,0.5,0.75$. For $W=0.05,0.15,0.25,0.5,0.75$, the fitted values of $(\gamma_c,a)$ are $(-0.02\pm0.06,4.6\pm0.9)$, $(-0.04\pm0.08,4.9\pm1.2)$, $(0.03\pm0.06,4.0\pm0.8)$, $(-0.04\pm0.07,4.9\pm0.9)$, and $(-0.15\pm0.06,5.9\pm0.8)$, respectively.
For $W=0$, we use the data from Fig.~2(b) of \cite{garciagarcia2026}, with $\gamma_c=0.00\pm 0.04$ and $a=2.55\pm 0.42$, and use a dashed line to distinguish these data from those obtained in this work. The error bars on the data points denote $95\%$ confidence intervals, whereas the uncertainties quoted for the fitted parameters correspond to one standard error ($1\sigma$).
		}\label{fig:QSD_gammaCfit}
\end{figure}

In Fig.~\ref{fig:QSD_gammaCfit}, we plot the correlation length for monitoring strengths $\gamma\in [0.5,0.85]$ and disorder strength $W\in [0,0.75]$. We use different disorder realizations for different values of disorder magnitude $W$, and for each fixed $W$ we use the same disorder realization across different quantum trajectories (with at least $10$ trajectories for each point). We have checked that the statistical errors obtained from trajectory-only averaging at fixed disorder and from averaging over both disorder realizations and trajectories differ only slightly.
Although the physical monitoring rate satisfies $\gamma\geq0$, we allow the fitted $\gamma_c$ to be slightly negative, since the disorder-induced replacement $\gamma\to\gamma+w$ effectively shifts the extrapolated singular point of the fitting form to $\gamma_c\simeq-w$. For $W\in[0.05,0.5]$, where $\gamma\gg w$, the shift is negligible and we fit directly with Eq.~(\ref{eq:lcorfit}). The resulting $|\gamma_c|<0.1$, even with relatively large uncertainties, provides conclusive evidence for the area-law phase at any finite $W$ and $\gamma$. For $W=0.75$, $w=W^2/3=0.1875$ becomes comparable to $\gamma$, so the shift is no longer negligible, and the value $\gamma_c=-0.15\pm0.06$ fitted using Eq.~(\ref{eq:lcorfit}) is close to $-w$.

Having demonstrated the absence of the MIPT, we now address the origin of the enhancement of the correlation length with disorder observed for different monitoring strengths in the weak disorder limit $W \in [0,0.5]$ in Fig.~\ref{fig:QSD_mu06} and Fig.~\ref{fig:QSD_gammaCfit}.  As a consequence of the symmetry change from class BDI at $W=0$ to class AIII in the presence of disorder, the correlation length increases substantially because the exponent in Eq.~(\ref{eq:RGAIII}) is twice that in Eq.~(\ref{eq:RGBDI}).
The symmetry-induced change in the exponential coefficient is most directly visible as a change in the slope of the semi-logarithmic plot in Fig.~\ref{fig:QSD_gammaCfit}. For $\gamma_c=0$, Eq.~(\ref{eq:lcorfit}) gives $\ln l_{\rm cor}=\ln b-\ln\gamma+a/\gamma$, so the prefactor $b/\gamma$ contributes only a subleading logarithmic correction, whereas the exponential term controls the dominant linear dependence on $1/\gamma$. To isolate this exponential contribution, we also fit the data to Eq.~(\ref{eq:lcorfit}) with $\gamma_c$ fixed to zero and define the fitted coefficient as $a_c\equiv a|_{\gamma_c=0}$. The corresponding values of $a_c$ for $W=0.05,0.15,0.25,0.5$, and $0.75$ are $4.3\pm0.7$, $4.4\pm1.0$, $4.4\pm1.1$, $4.3\pm1.2$, and $3.9\pm0.9$, respectively.
Indeed, we observe a large increase in $a_c$, from approximately $2.6$ for $W=0$ to $4.3\pm0.7$ for $W=0.05$, consistent with the expected factor-of-two change from $\frac{\sqrt{2}\pi}{2}$ to $\sqrt{2}\pi\simeq4.4$. For $W\in[0.05,0.5]$, the exponential coefficient $a_c$ remains close to $\sqrt{2}\pi$, while the correlation length increases monotonically with $W$.
This increase is likely originated by the fact the system does not yet fully feel the effect of disorder and therefore the change from the clean limit is gradual and not abrupt. In Sec.~\ref{app:NLSM2} of the Supplemental Material we provide a more precise explanation based on the RG-flow crossover from short length scales, where disorder is not yet relevant, to long length scales where it is.
At the larger disorder strength $W=0.75$, $l_{\rm cor}$ increases more slowly with decreasing $\gamma$ because $w=W^2/3$ becomes comparable to $\gamma$. In this regime, the same $\gamma\to\gamma+w$ effect also causes $l_{\rm cor}$ to decrease with increasing disorder at a fixed monitoring rate $\gamma$.

\begin{figure}[!htb]
	\centering
	\includegraphics[width=8cm]{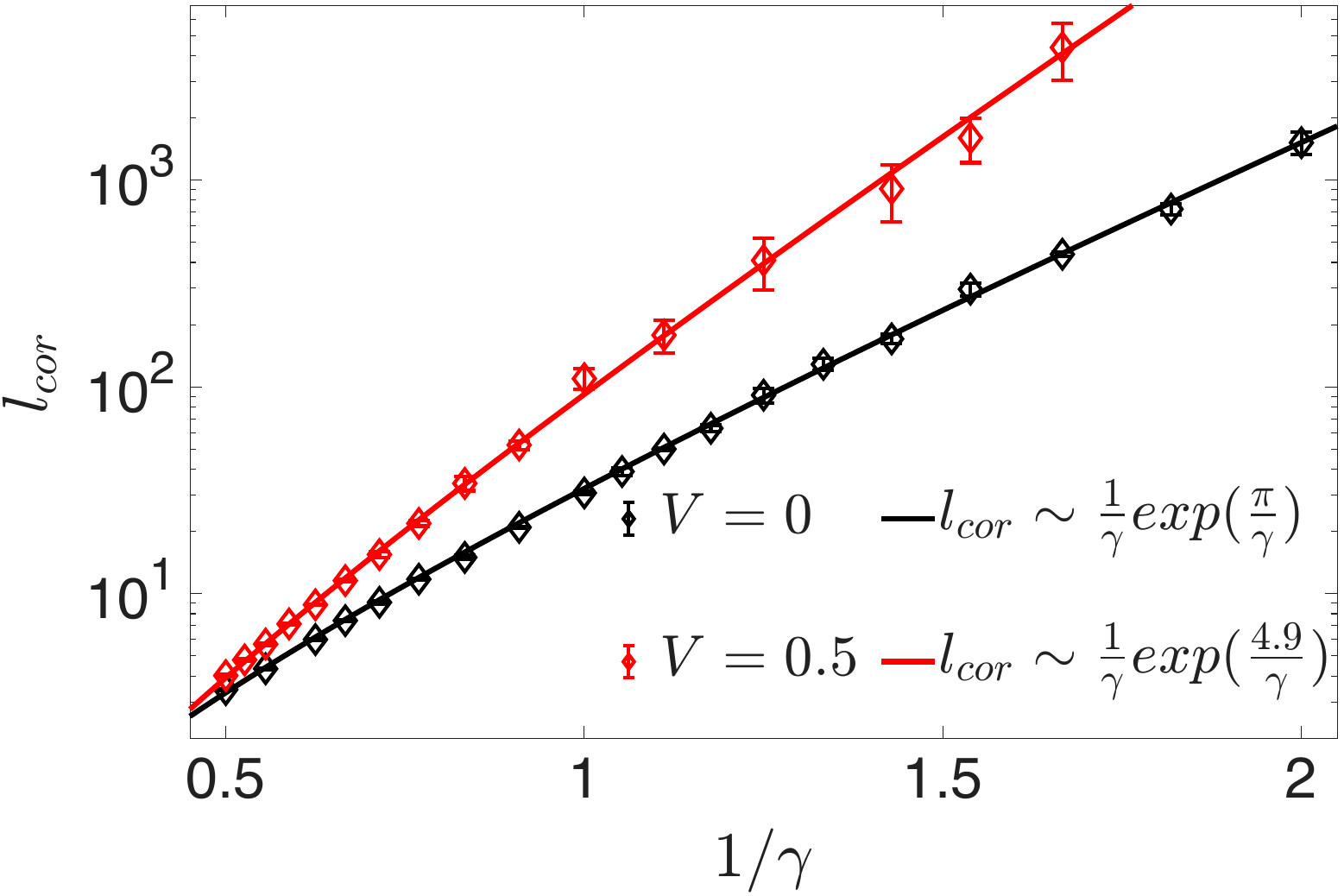}
	\caption{
		Correlation length $l_{\rm cor}$ as a function of the measurement strength $\gamma$ for the quasiperiodic system ($V=0.5$) under the PM protocol. The solid lines correspond to the analytical predictions Eqs.~(\ref{eq:RGBDI}) and~(\ref{eq:RGAIII}). For the quasiperiodic system, the numerical results are consistent with the AIII scaling Eq.~(\ref{eq:RGAIII}). When the monitoring strength $\gamma\geq0.8$, finite-size effects are negligible, so we focus on the smaller system size $L=2584$. When $\gamma\leq0.7$, we use the much larger system size $L=10946$.
		For comparison, data for the clean limit ($V=0$), class BDI,  are extracted from Fig.1(b) of Ref.~\cite{garciagarcia2026}, yielding fitting parameters $\gamma_c = 0.0 \pm 0.1$ and $a = \pi \pm 0.8$. For the quasiperiodic case ($V=0.5$), fitting to Eq.~(\ref{eq:lcorfit}) gives $\gamma_c = 0.0 \pm 0.03$ and $a = 4.9 \pm 0.6$. 
        }\label{fig:PM_fit_V0p5}
\end{figure}

We next study the quasiperiodic system under the PM protocol. Results for the correlation length are shown in Fig.~\ref{fig:PM_fit_V0p5}. Similar to the disordered case, the extracted correlation length displays exponential scaling with the inverse measurement rate $1/\gamma$. The fitting parameter $a = 4.9 \pm 0.6 $ when $V=0.5$ is consistent with the class AIII prediction in Eq.~(\ref{eq:RGAIII}). In contrast, for the clean limit $V=0$, we obtain $a = \pi \pm 0.8$, in agreement with the class BDI prediction. This comparison indicates that introducing the quasiperiodic potential changes the symmetry class and consequently modifies the universal scaling behavior. 
Consistently, the mutual information results shown in Fig.~\ref{fig:PM_I2_V0p5}, see section ~\ref{app:MI} of the Supplemental Material, suggest that the critical measurement rate $\gamma_c$ vanishes in the thermodynamic limit.

\noindent {\it Analytical Entanglement Dynamics.-}
Finally, for the disordered case under the QSD protocol, we present the RG-flow prediction for the correlation length, which provides the main analytical demonstration of the absence of the MIPT, as detailed in Secs.~\ref{app:NLSM1} and~\ref{app:NLSM2} of the Supplemental Material.
The steady-state entanglement dynamics is governed by the one-loop RG equation for the $SU(R)$ NLSM in the replica limit $R\to1$,
\begin{equation}\label{eq:disorder_oneloop}
	\frac{\partial D}{\partial \log l}=-\frac{1}{4\pi}+O(1/D),\quad D(l_0)=\frac{\sqrt{2}}{4(\gamma+w)},
\end{equation}
where $D(l)$ is the running NLSM stiffness at the RG length scale $l$, and $l_0=\frac{\sqrt{2}}{2(\gamma+w)}$ is the initial RG length scale. Since the right-hand side is still negative, the EE is in the area law phase for any finite $w$ and $\gamma$. Its solution gives the correlation length as a function of $w$ and $\gamma$,
\begin{equation}\label{eq:lcorW}
	l_{\mathrm{cor}} \sim l_0\exp\left(\frac{\sqrt{2}\pi}{\gamma+w}\right).
\end{equation}
Therefore, disorder modifies the clean-limit correlation length in Eq.~(\ref{eq:RGBDI}) in two respects:\\
(1). It changes the symmetry class from BDI to AIII, so for weak disorder, the correlation length given by Eq.~(\ref{eq:RGAIII}) is larger than that in the clean limit.\\
(2). It replaces the monitoring strength $\gamma$ with $\gamma+w$, which reduces the correlation length.

For our main numerical results with $W\in[0.05,0.5]$, where $\gamma\gg w$, effect~(2) is subleading, and the correlation-length scaling is governed primarily by the symmetry-changing effect~(1), consistent with the AIII exponential coefficient extracted from the fits. For $W=0.75$, where $\gamma/w\sim O(1)$, effect~(2) becomes important and accounts for both the slower increase of the correlation length with decreasing $\gamma$ than in the $\gamma\gg w$ AIII regime and the fitted shift $\gamma_c\simeq-w$.
 
 {\it Conclusions.-} 
 We have studied the entanglement dynamics of monitored 1d non-interacting $U(1)$ fermions in the presence of a disordered or a quasiperiodic potential using quantum state diffusion and projective measurement protocols respectively. In both cases, we have shown that the EE after saturation does not scale with the system size for any finite monitoring rate, independently of the strength of the disordered or quasiperiodic potential, so no MIPT occurs in the system. Our conclusion is supported by large-scale simulations reaching $L\simeq18000$, much larger than in previous studies. The simulations employ CUDA/C++ and MATLAB implementations together with a finite-size scaling analysis.
 
 In the case of a disordered potential, the absence of the MIPT is also shown by mapping the entanglement dynamics onto a NLSM and using field theory techniques. Another outcome of this analytical approach, supported by the numerical results, is that in the weak-disorder and weak-monitoring regime $w \ll \gamma \ll 1$, the correlation length increases with disorder and grows exponentially as the monitoring strength decreases. It is also larger than in the clean limit because the symmetry class is different. This enhancement justifies the need to reach $L\simeq18000$ to demonstrate numerically the lack of a MIPT. In addition, when both disorder and monitoring are weak but their ratio satisfies $w/\gamma\sim O(1)$, the correlation length decreases as the disorder strength increases.
\section{End Matter}
In this End Matter, we first explain how the EE can be computed numerically, either directly from the eigenvalues of the correlation matrix or, within the leading-cumulant approximation, from the density correlation function. We then briefly outline the replicated Keldysh treatment of entanglement dynamics in the disordered monitored system and trace the symmetry structure from the closed-system Hamiltonian, through the collective-field action and saddle-point manifold, to the NLSM target space.

{\it Numerical computation of the entanglement entropy.-}
The full system we investigate remains quadratic even in the presence of monitoring. Wick’s theorem therefore enables us to express any observable in terms of the $L\times L$ correlation matrix $D$, defined by $D_{ij}(t)\equiv\langle\psi(t)|c_j^\dagger c_i|\psi(t)\rangle$. The explicit time evolution of $|\psi(t)\rangle$ under each protocol is given in Appendix~\ref{app:protocol}.

The EE is defined as $S_A(t)=-\Tr_A(\rho(t)\log(\rho(t)))$. Unless stated otherwise, the subsystem is $A=\{1,2,\cdots,L/2\}$, $\rho(t)=|\psi(t)\rangle\langle\psi(t)|$ is the density matrix, and the EE is expressed as
$S_A(t)=-\sum_{i=1}^{L/2}\left(\lambda_i(t)\log(\lambda_i(t))+(1-\lambda_i(t))\log(1-\lambda_i(t))\right)$,
where $\lambda_i$ are the eigenvalues of $D_{i_Aj_A}$ with $i_A,j_A\in A$. The system-size scaling of the EE after saturation is an indicator of a MIPT.

However, it is more convenient to expand the EE as a series of cumulants of the occupation number. Perturbatively, it was found that only the second cumulant, which is proportional to the density-density correlation function $C(x-y,t)$, needs to be considered~\cite{levitov2009,poboiko2023},
\begin{equation}
S_A(t)\approx\frac{\pi^2}{3}\int_0^{L/2}dx\int_0^{L/2}dy\,C(x-y,t).
\label{eq:C2}
\end{equation}
For computational purposes, we use $C(r,t)=\overline{C(|x-y|=r,t)}$, where the overline denotes an average over all pairs $(x,y)$ satisfying $r=|x-y|$, as well as over quantum trajectories, namely, different realizations of $|\psi(t)\rangle$ generated by the monitored dynamics. On a lattice, $C_{ij}(t)$, with $i,j$ denoting site indices, is obtained from the correlation matrix $D_{ij}(t)$ using Wick’s theorem: $C_{ij}(t)=-D_{ij}(t)D_{ji}(t)$ and $C(r,t)=\overline{C_{ij}(t)}$, where we consider only $i\neq j$ with $|i-j|=r$. We focus on long times $t\geq L/2$, when the EE has reached its saturation value. We have checked that, up to an overall prefactor, Eq.~(\ref{eq:C2}) remains accurate~\cite{poboiko2023,poboiko2023a} beyond the perturbative regime $\gamma\ll1$.

{\it Keldysh formalism for entanglement dynamics.-}
We set up the Keldysh formalism for the Hamiltonian in Eq.~(\ref{eq:Dis_Ham}) under quantum-state-diffusion monitoring, whose state evolution is described by a stochastic Schrödinger equation (SSE). We consider the monitored system in the long-time limit, where the system is fully scrambled and the ensemble-averaged density matrix becomes maximally mixed within the fixed-particle-number sector, corresponding to infinite temperature. We further restrict the system to half filling, $N=L/2$, which simplifies the following analysis of the causal structure in Keldysh space.

For a general cumulant of order $m$, $\overline{\langle(\delta N_A)^m\rangle_c}$ of the particle number $N_A=\sum_{x\in A}n(x)$ in a subregion $A$, where $\langle\cdots\rangle=\Tr(\cdots\rho)$, $\overline{\cdots}$ denotes the trajectory average, and the subscript $c$ denotes the connected part, the observable is nonlinear in the density matrix $\rho$ and therefore cannot be described by the standard master equation for the single-copy density matrix. Instead, we consider $m$ copies of the system, formulate the replicated SSE, and construct the master equation for the replicated density matrix $\rho^m\equiv\overline{\rho^{\otimes m}}$. Disorder and monitoring-induced It\^{o} randomness then generate couplings among the bra-bra, ket-ket, and bra-ket sectors, producing additional inter-replica interaction channels absent from the standard single-copy master equation.

Continuous monitoring measures the local particle densities $n_i$, with $i=1,2,\ldots,L$ labeling the sites, and we employ the normalized SSE. However, in the numerical simulation, as detailed in Section~\ref{app:protocol} of the Supplemental Material, the normalization factor accumulated over a finite evolution interval remains non-negligible in the continuum limit $dt\to0$. We therefore perform a QR decomposition at each time step $t\to t+dt$, and the corresponding normalization factor can be interpreted as a Born-like trajectory weight. The extra normalization factor is canceled in the path-integral formalism by introducing $R-m$ additional replicas with $R>m$, followed by an analytic continuation to the replica limit $R\to1$~\cite{poboiko2023}. This analytic continuation is allowed by the analytic dependence of the master equation for $\rho^R$ on $R$. The replica limit $R\to1$ also selects the replica-symmetric sector in which the causal saddle-point equation closes within the chosen ansatz, thereby simplifying the saddle-point analysis.

We construct the path integral from the master equation for the replicated density matrix $\rho^R$. To be explicit, in the spatial continuum limit $i\to x$, we map the fermionic operators $c_i,c_i^\dagger$ to the Grassmann fields $\psi(x),\bar{\psi}(x)$ using symmetrized (Weyl) ordering, and associate the ket and bra with the Keldysh $\pm$ indices, respectively. We further introduce the compact fields $(\Psi,\bar{\Psi})$ in replica$\times$Keldysh space by defining $\vec{\psi}=(\psi^{(1)},\psi^{(2)},\cdots,\psi^{(R)})$, $\vec{\bar{\psi}}=(\bar{\psi}^{(1)},\bar{\psi}^{(2)},\cdots,\bar{\psi}^{(R)})^T$, with the parenthesized superscripts labeling replicas, and $\Psi=(\vec{\psi}^+,\vec{\psi}^-)^T$, $\bar{\Psi}=(\vec{\bar{\psi}}^+,\vec{\bar{\psi}}^-)$, so that the action is written as $iS[\bar{\Psi},\Psi]$. We then introduce the collective fields $\mathcal{G}$ and $Q$ through the identity $1=\int D\mathcal{G}DQ\,\exp\Big(i\Tr\left[Q(\mathcal{G}-i\Psi\bar{\Psi})\right]\Big)$ and integrate out $(\Psi,\bar{\Psi})$ to obtain the action $iS[\mathcal{G},Q]$. Next, we perform the Gaussian integration over $\mathcal{G}$ in the independent channels to obtain the effective action $iS[Q]$. For channels with positive Gaussian coefficients, the divergent Gaussian integral is defined by analytically continuing the quadratic kernel and the independent components of $\mathcal{G}$, ending on the imaginary-axis configuration $\mathcal{G}=iX$, as justified within our formalism below Eq.~(\ref{eq:SeparateChannel}) in Section~\ref{app:NLSM1} of the Supplemental Material.

In the saddle-point analysis, we choose the replica-symmetric configuration at $R=1$, and for the temporal bilocal field $Q(t,t')$, we focus on the equal-time saddle at $t=t'$. We perform the Keldysh rotation to establish the causal structure and obtain the standard saddle $\Lambda=\sigma_K^z$, where the subscript $K$ denotes the $2\times2$ Keldysh space. We then perform a slow-field expansion around the saddle using $Q=T\Lambda T^{-1}$, with $T=e^{i\Phi\sigma_K^x/2}$, and integrate out the massive $t\neq t'$ sector together with the massive rotation-generator directions proportional to $\sigma_K^{y,z}$. At this order, these massive modes generate only finite local matching corrections to the NLSM parameters, while the infrared RG flow is governed by the remaining soft modes. The soft modes are parameterized by $U=e^{i\Phi}$, yielding the standard NLSM $iS^K[U]=-D/2\Tr\sum_{\mu=\tilde{x},\tilde{t}}(\partial_\mu U^\dagger)(\partial_\mu U)$, where the superscript $K$ indicates that the action is written in the Keldysh-rotated basis. Here $\tilde{t}=v_*t$, $\tilde{x}=x$, and $v_*=\sqrt{2}$ define the rescaled coordinates for which the temporal and spatial gradient terms take the same isotropic form, and the NLSM stiffness is $D=1/(2\sqrt{2}(\gamma+w))$.

{\it Symmetry classification from Hamiltonian to NLSM.-}
In the absence of disorder, the clean Hamiltonian in Eq.~(\ref{eq:Dis_Ham}) with $V_i=0$ possesses both time-reversal symmetry (TRS) and chiral symmetry (CS). A disordered potential explicitly breaks CS, while a fixed realization of the monitoring dynamics breaks TRS. In the clean monitored problem, the remaining CS is lifted to the PHS$^\dagger$ constraint of the doubled Lagrangian operator.

The averaged replicated Keldysh action $iS[\Psi,\bar{\Psi}]$ contains the two channels $(\bar{\Psi}\Psi)$ and $(\bar{\Psi}\sigma_K^z\Psi)$, and we consider transformations $\Psi\to \hat{U}\Psi$ that leave the action invariant. In the disordered case, $\hat{U}$ takes values in $U(R)\times U(R)$, while in the clean case, the additional symmetry constraint inherited from the clean Hamiltonian imposes a reality condition that restricts $\hat{U}$ to $O(R)\times O(R)$.
Embedding these transformations in the full $2R$-dimensional replica$\times$Keldysh space, we can associate them with the AZ classifying spaces AIII, $U(2R)/[U(R)\times U(R)]$, and BDI, $O(2R)/[O(R)\times O(R)]$, respectively. The physical meaning of these classifying spaces can also be understood in terms of the doubled Lagrangian operator for a single trajectory~\cite{poboiko2024}. To avoid confusion, we emphasize that the symmetry groups of the averaged replicated action itself are the direct-product groups $U(R)\times U(R)$ and $O(R)\times O(R)$.

The Goldstone modes parameterize the saddle manifold $G/H$, where $G$ is the full continuous symmetry group generating the saddle configurations and $H\subset G$ is the stabilizer subgroup that leaves the reference saddle invariant. For the disordered and clean cases, the target spaces are $SU(R)$ and $SU(2R)/Sp(2R)$, respectively. Here we remove the $U(1)$ contribution from the $R=1$ replica-symmetric sector, since this contribution decays to zero in the long-time limit and therefore does not contribute to the steady-state density correlation function $C(r)$, which is the main observable studied in the main text. Therefore, adding disorder changes the NLSM target space from $SU(2R)/Sp(2R)$ to $SU(R)$, leading to a different one-loop renormalization of the stiffness $D$. Using the background-field method, we obtain the RG flow from the Ricci tensor of the target space and the factor-of-two difference in the correlation-length exponential coefficient between the clean and disordered cases. We note the similarity to Anderson localization~\cite{evers08} in the mapping from the Hamiltonian symmetry class to the NLSM target space, although for the monitored system, deriving the standard NLSM form from the microscopic dynamics is highly nontrivial.

{\it Note added.-} While finalizing this manuscript, we became aware of the recent preprint \cite{liao2026measurement} which also studies the role of disorder in the entanglement dynamics of monitored free fermions using similar techniques. It agrees with our main conclusion regarding the absence of a MIPT in one dimension. However, contrary to our analytical and numerical results, in the weak disorder limit, it predicts a monotonic decrease of the correlation length with disorder. Moreover, numerically, the maximum system size in Ref. \cite{liao2026measurement} is $L\simeq 2000$ while our numerical simulations reach $L\simeq 18000$.

\acknowledgments

A. M. G. thanks Yan Fyodorov, Kohei Kawabata and Marco Schiro for illuminating discussions. B. F. thanks Nils Mantik, Igor Poboiko, Igor Gornyi and Alexander Mirlin for valuable comments on the unpublished numerical results, especially for bringing to my attention that a random potential should change the symmetry class from BDI to AIII \cite{Mantikupblished}. C.Y. thanks Nils Mantik for interesting discussions. We thank Nils Mantik, Igor Poboiko, Igor Gornyi and Alexander D. Mirlin for insightful comments on the first version of the manuscript posted in arxiv, especially for indicating an error in the treatment of the NLSM that we could subsequently identify and fix. We thank Zhenyu Xiao for interesting conversations. We acknowledge support from the National Natural Science Foundation of China (NSFC): Individual Grant No. 12374138. B.F. acknowledges support from the China Postdoctoral Science Foundation (Grant numbers: 2023M732256, 2023T160409, GZB20230420) and from the Deutsche Forschungsgemeinschaft (DFG, German Research Foundation) – 553096561. C.Y. acknowledges support from a T.D. Lee scholarship.
\bibliography{quasiranmonitoring.bib}

\clearpage
\onecolumngrid
\setcounter{secnumdepth}{2}
\setcounter{section}{0}
\setcounter{subsection}{0}
\setcounter{equation}{0}
\setcounter{figure}{0}
\setcounter{table}{0}

\renewcommand{\thesection}{S\arabic{section}}
\renewcommand{\thesubsection}{S\arabic{section}.\arabic{subsection}}
\renewcommand{\theequation}{S\arabic{equation}}
\renewcommand{\thefigure}{S\arabic{figure}}
\renewcommand{\thetable}{S\arabic{table}}

\begin{center}
{\large\bfseries Supplemental Materials\par}
\end{center}
This Supplemental Material provides technical details and additional numerical results supporting the analysis presented in the main text. Section~\ref{app:protocol} describes the numerical implementation of the projective measurement (PM) and quantum state diffusion (QSD) protocols. Section~\ref{app:MI} presents numerical results for the mutual information in the quasiperiodic model, showing that the apparent finite-size crossing disappears for sufficiently large system sizes, $L\geq 2500$, and thereby confirming the absence of a MIPT in the thermodynamic limit.

Section~\ref{app:NLSM1} derives the replicated master equation for the continuously monitored system with static disorder, including the $R\to1$ replica construction that cancels the trajectory-normalization factor, and maps it to the collective-field Keldysh action in terms of $\mathcal{G}$ and $Q$. After performing the channel-resolved Gaussian integration over $\mathcal{G}$, we determine the homogeneous half-filled replica-symmetric causal saddle in the equal-time sector at $R=1$.

Section~\ref{app:NLSM2} traces the symmetry structure from the Hamiltonian, through the replicated action and the fixed-trajectory AZ classifying spaces, to the saddle manifold, distinguishing between the disordered AIII and clean BDI cases. We separate the replica-symmetric $U(1)$ mode from the $SU(R)$ replicon and restrict the theory to the gapless fluctuations around the saddle. The effects of the various massive sectors enter only through the short-distance matching of the NLSM parameters, and the resulting gapless theory is the $SU(R)$ NLSM. We use the background-field method to compare its RG flow with that of the clean $SU(2R)/Sp(2R)$ theory, determine the correlation-length scaling across the disorder crossover, and connect the replicon NLSM to the physical density correlation function $C(r)$.

\section{PM and QSD protocols}\label{app:protocol}
This section introduces the numerical details of the measurement protocols. 
For the PM protocol, we initialize the system in the N\'{e}el state $|\psi(t=0)\rangle = |10101010\cdots\rangle$ and evolve the monitored system using the corresponding correlation matrix $D_{ij}(t) = \langle \psi(t)| c_j^\dagger c_i |\psi(t)\rangle $. For numerical stability and efficiency, the time evolution is performed directly on this correlation matrix. Between two successive measurements, the system evolves unitarily under the Hamiltonian $H$. The updated correlation matrix is given by $D (t+\tau) = e^{-iH\tau} D(t) e^{iH\tau} $, where $\tau = -{\rm ln}(\eta)/(\gamma N)$ and $\eta \in (0,1]$ is a uniformly distributed random number. $N=L/2$ is the particle number and $\gamma$ is the measuring rate per site.
At each measurement step, namely, at time $t+\tau$, we first randomly select a site $j$, and compute the local occupation number $p_j = D_{j,j}$. We then generate a random value $p_c \in[0,1]$. If $p_j\ge p_c$, we perform the projection using the operator $ \hat{P}_1(j) = \hat{n}_j $. The corresponding correlation matrix is updated as follows:
\begin{equation}
	D'_{i,i'} = \langle \psi' | c_{i'}^\dagger c_i | \psi' \rangle
	= \begin{cases}
		1, ~i=i'=j \\
		0,  (i=j, i\ne i') ~\text{or}~ (i'=j, i\ne i') \\
		D_{i,i'} - \dfrac{D_{i,j}D_{j,i'}}{D_{j,j}}, ~ \text{\rm otherwise}.
	\end{cases}
	\label{eq:QJ_update}
\end{equation}

If $p_j< p_c$, we perform the projection operation $ \hat{P}_0(j) = 1 - \hat{n}_j $. The corresponding updated correlation matrix elements are given by:
\begin{equation}
	D'_{i,i'} = \langle \psi' | c_{i'}^\dagger c_i | \psi' \rangle
	= \begin{cases}
		0, ~~~~~ (i=j) ~\text{or}~ (i'=j) \\
		D_{i,i'} + \dfrac{D_{i,j}D_{j,i'}}{1-D_{j,j}}, ~ \text{otherwise}
	\end{cases}
	\label{eq:PM_update_0}
\end{equation}

After each measurement, we reconstruct the new orthonormal state $|\psi(t+\tau)\rangle$ from the correlation matrix $D$ using the thin singular value decomposition. The new state matrix $ U $ characterizes the new state $ |\psi(t+\tau) \rangle $ at time $ t + \tau $.
We repeat the calculation until $t \geq L/2$ to ensure that the system reaches the steady state at which the EE does not experience any net growth.

In the QSD protocol, we likewise initialize the system in the N\'{e}el state $|10101010\cdots\rangle$ and evolve the coefficient matrix $U$ of the wave function in the particle-number basis, which is defined as:
\begin{equation}
	|\psi(t)\rangle = \prod_{k=1}^{N} \left[\sum_{j=1}^L U_{jk}(t) c_j^\dagger\right] | \text{vac} \rangle
\end{equation}
where $|\text{vac} \rangle$ is the vacuum state annihilated by any $c_i, i=1,2\cdots, L$; $U$ is an $L\times N$ matrix satisfying $U^\dag U=\mathbf{1}_{N}$, where $\mathbf{1}_{N}$ denotes the $N\times N$ identity matrix.
Unlike the PM protocol, in which measurement events are performed randomly in space and time, here we continuously measure the system uniformly in space and time by introducing independent Gaussian random noise $dW_i^t$ with zero mean at each site $i$. The increments satisfy $dW_i^t dW_j^t=\gamma\delta_{ij}dt$, where $\gamma$ is the measurement rate. The measurement term $i\sum_i dW_i^t\hat n_i$ is non-Hermitian, so that the system evolves non-unitarily according to the state-evolution equation:
\begin{equation}
U(t+dt) \propto \exp\biggl(-i\tilde{H} dt+d W^t +(2\langle \hat{n}\rangle_t-\mathbf{1}_{L})\gamma dt\biggr) U(t)
	\label{eq:QSD_evo}
\end{equation}
where $\tilde{H}_{ij}=J\delta_{i\pm1,j}+V_i\delta_{ij}$ is the $L\times L$ coefficient matrix of the Hamiltonian $H$, namely, $H=\sum_{ij}\tilde{H}_{ij}c_i^\dag c_j$. The matrices $dW^t$ and $\langle\hat n\rangle_t$ are $L\times L$ diagonal matrices, with $i$-th diagonal elements $dW_i^t$ and $\langle\hat n_i\rangle_t=\sum_jU_{ij}U_{ij}^*$, respectively.
In practice, we use the fourth-order Runge–Kutta method to evolve the matrix $U(t)$ with a time step of $dt=0.05$, and we have verified that for $dt \le 0.1$, the results are identical to those obtained by exact diagonalization. After each time step $t\rightarrow t+dt$, we perform a QR decomposition of the matrix $U$, writing $U=QR$, where $Q$ has orthonormal columns satisfying $Q^\dagger Q=\mathbf{1}_N$ and $R$ is an upper-triangular matrix. We then set $U=Q$. As in the PM protocol, we repeat the calculation until $t \geq L/2$ when the system reaches a steady state. For $t \ge L/2$, we compute the density–correlation function and the entanglement entropy from the correlation matrix $D(t)=U(t)U(t)^\dag$. On a single A100, evolving one trajectory over $t\in [0,L]$ requires approximately 11 hours for $L=8192$.

\section{Mutual information $\mathcal{I}_2$} \label{app:MI}
\begin{figure}[!htb]
	\centering
	\includegraphics[width=8cm]{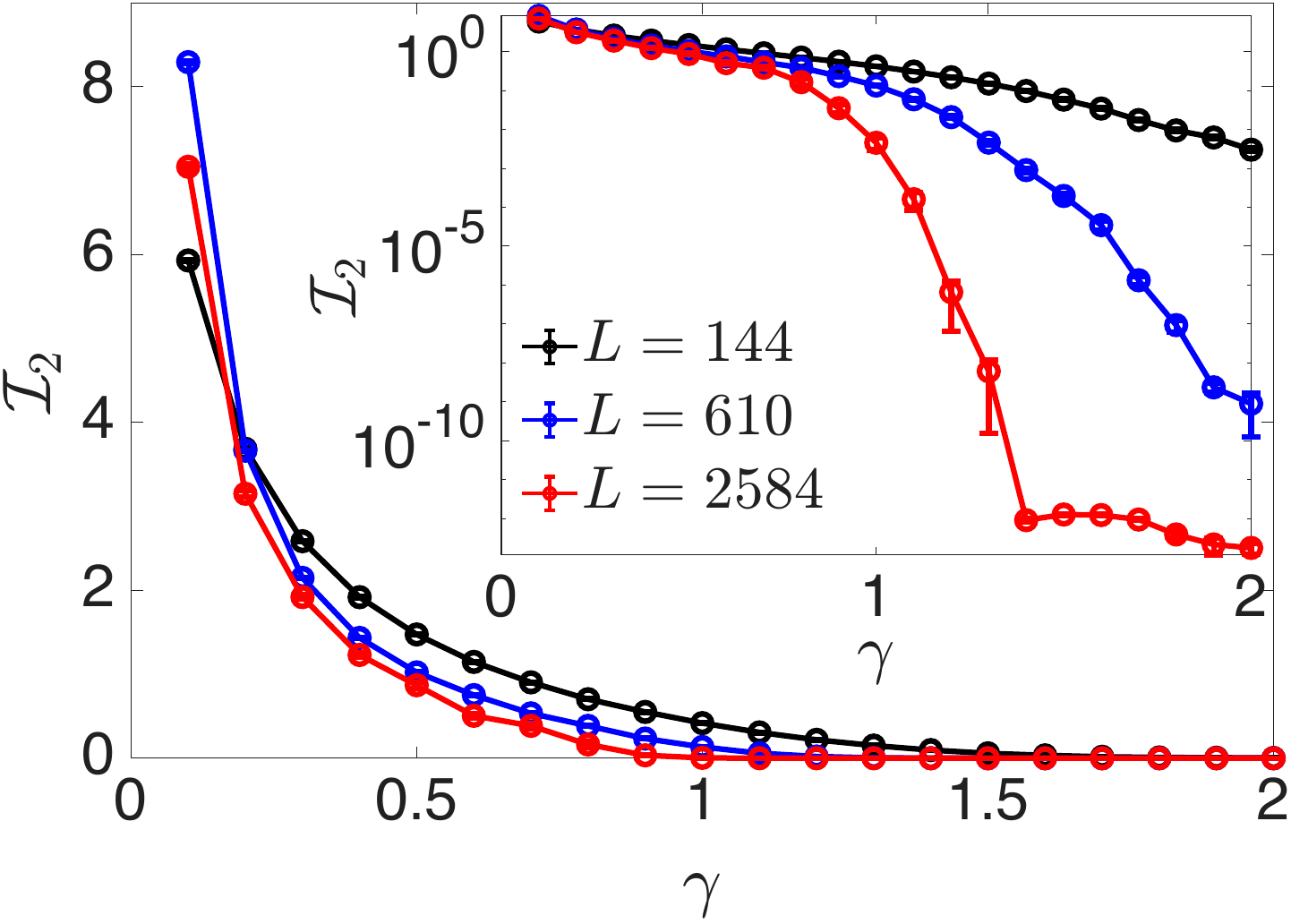}
	\caption{
    The mutual information $\mathcal{I}_2$ as a function of measurement strength $\gamma$ for the quasiperiodic system ($V = 0.5$) under the projective measurement. For small system sizes, the curves exhibit an apparent crossing point around $\gamma \sim 0.2$, which could be mistaken as a signature of a critical transition. However, this crossing is a finite-size effect, consistent with previous studies \cite{matsubara2025}. As the system size increases, the crossing gradually disappears. In particular, for sufficiently large systems  $L \ge 2584$, no crossing is observed with smaller sizes such as $L=610$. This indicates the absence of a true phase transition in the thermodynamic limit. 
	}\label{fig:PM_I2_V0p5}
\end{figure}

In Fig.~\ref{fig:PM_I2_V0p5}, we present results for the mutual information $\mathcal{I}_2$ for the quasiperiodic model that confirm the absence of a MIPT.  

\section{Keldysh field theory and the saddle-point analysis} \label{app:NLSM1}
In the quantum state diffusion (QSD) protocol, the Lindblad master equation is unraveled into stochastic pure-state trajectories by introducing independent real Itô increments $dW_i^t$, $i=1,2\cdots L$, which satisfy $\mathbb{E}[dW_i^t]=0$ and $dW_i^t dW_j^t=\gamma\,\delta_{ij}\,dt$. In the presence of static on-site disorder $V_i$, the corresponding normalized stochastic Schrödinger equation (SSE) contains both the stochastic measurement term and the deterministic nonlinear Itô drift $-\frac{\gamma}{2}(\hat{n}_i-\langle\hat{n}_i\rangle_t)^2dt$, where $\langle\hat{n}_i\rangle_t=\langle\psi_t|\hat{n}_i|\psi_t\rangle$. This nonlinear drift ensures the normalization of each quantum trajectory:
\begin{equation}
d|\psi_t\rangle=-i \hat{H}_0 dt |\psi_t\rangle +\sum_{i=1}^L \Big( (\hat{n}_i-\langle \hat{n}_i \rangle) dW_i^t
-i \hat{n}_i V_i dt -\frac{1}{2} (\hat{n}_i-\langle \hat{n}_i\rangle_t)^2\gamma dt\Big)|\psi_t\rangle
\label{eq:SSE}
\end{equation}
where $\hat{H}_0=J\sum_{i=1}^L(\hat{c}^\dagger_{i}\hat{c}_{i+1}+\hat{c}^\dagger_{i+1}\hat{c}_{i})$ is the clean nearest-neighbor free-fermion Hamiltonian, and $V_i$ is uniformly distributed in $[-W,W]$, as defined in Eq.~(\ref{eq:Dis_Ham}). In the perturbative regime of weak disorder, up to $O(W^2)$, the effect of $V_i$ is equivalent to that of a Gaussian-distributed random variable $\tilde{V}_i$ with zero mean, variance $W^2/3$, and $\langle \tilde{V}_i\tilde{V}_j\rangle=0$ for $i\neq j$. This follows from $\mathbb{E}[e^{V_i O}]=\frac{\sinh(WO)}{WO}=1+\frac{(WO)^2}{6}+O(W^4)$, where $O$ denotes a general variable. In the following analysis, we take the Gaussian variable $\tilde{V_i}$ instead of $V_i$ for simplicity and fix $J=1$.

We then construct the stochastic evolution of the density operator by combining the bra and ket evolutions, $d\rho_t=|d\psi_t\rangle\langle\psi_t|+|\psi_t\rangle\langle d\psi_t|+|d\psi_t\rangle\langle d\psi_t|$. Applying the It\^{o} product rule gives
\begin{equation}
d\rho_t=-i\left[\hat H_0+\sum_{i=1}^L \tilde{V}_i\hat n_i,\rho_t\right]dt+\sum_{i=1}^L\left\{\hat n_i-\langle\hat n_i\rangle_t,\rho_t\right\}dW_i^t-\frac{\gamma}{2}\sum_{i=1}^L\left[\hat n_i,\left[\hat n_i,\rho_t\right]\right]dt.
\label{eq:rho1}
\end{equation}
Here, the standard double-commutator term is obtained by combining the deterministic It\^{o} drift in the SSE with the quadratic variation $dW_i^tdW_j^t=\gamma\delta_{ij}dt$. This It\^{o} rule reflects the self-averaging quadratic variation of Brownian increments in the continuous-time limit and does not itself constitute an average over quantum trajectories. We then average over both quantum trajectories and static disorder, $d\rho_t\to\overline{d\rho_t}$. Under the trajectory average, the remaining stochastic term $\sum_i\{\hat n_i-\langle\hat n_i\rangle_t,\rho_t\}dW_i^t$ vanishes because the Wiener increments have zero conditional mean. We assume that the random seeds used for the trajectories and disorder realizations are independent, so that the order of the two integrations can be interchanged.
In the present real-time setup, averaging over disorder can be tricky because static disorder generally correlates all time points, so that in many cases, especially for interacting systems, we cannot write down a Markovian master equation. Here, the stochastic Schr\"odinger equation in Eq.~(\ref{eq:SSE}) contains only quadratic terms, so that the infinite Dyson series can be factorized into two-point terms that are bilocal in time and expressed in a master-equation form,
\begin{equation}
\overline{\mathcal T\exp[-i\sum_{i=1}^L\tilde{V}_i\int_0^Tdt\mathcal L_i^d(t)]}
=
\mathcal T\exp[-\frac12\sum_{i,j=1}^L\overline{\tilde{V}_i \tilde{V}_j}\int_0^Tdt\int_0^Tdt'\mathcal L_i^d(t)\mathcal L_j^d(t')]
\label{eq:dis_ave}
\end{equation}
where $\mathcal L_i^d(t)\rho=[(\hat n_i)^I(t),\rho]$, $T$ is the total time range. Wick's theorem is justified because the disorder $\tilde V_i$ is Gaussian and the superoperator $\mathcal L$ depends linearly on the density matrix.
The superscript $I$ denotes the interaction picture with respect to $\hat H_0$, introduced only to isolate the static-disorder contribution. After performing the disorder average, we return to the full evolution and suppress the superscript $I$ below. As a result, $\overline{\mathcal T\exp[\sum_{i=1}^L\int_0^T(\tilde{V}_iO_i(t)dt+dW_i^tQ_i(t))]}=\mathcal T\exp[\frac12\sum_{i,j=1}^L\int_{[0,T]^2}(\overline{\tilde{V}_i\tilde{V}_j}O_i(t)O_j(t')dtdt'+\overline{dW_i^tdW_j^{t'}}Q_i(t)Q_j(t'))]$ follows from Gaussian integration over the two independent measures for $\tilde{V}_i$ and $dW_i^t$. At leading order, static disorder gives the additional term $-\frac{W^2}{6}\int_{[0,T]^2}dtdt'\sum_{i=1}^L[\hat n_i(t),[\hat n_i(t'),\rho]]$. The resulting non-Markovian master equation can be written as 
\begin{equation}
\overline{d\rho_t}=-i[\hat H_0,\overline{\rho_t}]dt-dt\int_0^t dt'\sum_{i=1}^L(\frac{W^2}{3}+\gamma\delta(t-t'))[\hat n_i(t),[\hat n_i(t'),\overline{\rho_t}]].
\label{eq:rhoave}
\end{equation}
Here we adopt the endpoint convention $\int_0^t dt'\delta(t-t')=1/2$, which follows from a symmetric regularization of the delta function, since only half of the regularized peak lies within the integration interval.

Our quantity of interest is the entanglement dynamics like entanglement entropy $-\Tr_A(\rho\log(\rho))$, where $A$ is the subregion. Because the initial state is Gaussian and the particle-number-conserving quadratic Hamiltonian and density-monitoring dynamics preserve Gaussianity along each quantum trajectory, the trajectory entanglement entropy can be expressed in terms of the cumulants of the particle-number operator $\sum_{x\in A}\hat{n}(x)$, where $A$ denotes the subsystem under consideration. These cumulants are nonlinear functionals of the trajectory density matrix and are represented in the replica formalism by quantities of the form $\Tr(\cdots\rho^{\otimes m})$, where $\rho$ is the normalized density matrix of an individual trajectory. Generic measurements naturally produce an unnormalized trajectory density matrix $\tilde{\rho}$, with the corresponding normalized state given by $\rho=\tilde{\rho}/\Tr(\tilde{\rho})$. Under the Born rule, each trajectory carries the probability weight $\Tr(\tilde{\rho})$. For quantities linear in $\rho$, this weight cancels the normalization denominator, so their trajectory averages depend linearly on $\tilde{\rho}$ and are governed by the master equation. For nonlinear quantities such as $\Tr(\cdots\rho^{\otimes m})$, the cancellation is incomplete and leaves the factor $[\Tr(\tilde{\rho})]^{1-m}$. Following Ref.~\cite{poboiko2023}, this factor is represented by introducing an additional $R-m$ replicas, indexed by $m+1,m+2,\ldots,R$, and subsequently taking the replica limit $R\to1$. The required analytic continuation is explicit in the resulting replicated master equation.

We note that the density matrix in Eq.~(\ref{eq:rho1}) is normalized, with the measurement record coupled to the local density deviation $\hat n_i-\langle\hat n_i\rangle_t$. In the numerical evolution with a finite time step $dt$, however, the stochastic update in Eq.~(\ref{eq:QSD_evo}) produces an $L\times N$ matrix $U$ representing the wave-function, which must be normalized through the QR decomposition. This is an intrinsic feature of the finite-step nonunitary measurement evolution rather than a numerical artifact, since omitting the QR normalization leads to an accumulated norm drift over any finite evolution interval. The norm encoded in the triangular factor of $U=QR$ can therefore be interpreted as the finite-step counterpart of the Born likelihood, providing the same physical basis for the replica limit $R\to1$. A closely related construction was used in Ref.~\cite{poboiko2025a}, where continuous quantum diffusion was represented by finite-step weak-measurement Kraus operators and the same replica trick was applied.

We therefore work in the $R$-replicated Hilbert space $\mathcal H^{\otimes R}$. The same realizations of the monitoring noise $dW_i^t$ and static disorder $\tilde{V}_i$ are shared by all $R$ copies, while the clean Hamiltonian $\hat H_0$ acts identically in each copy. For any operator $\hat A$, we define its action in replica $a$ as $\hat A^{(a)}=\mathbb I^{\otimes(a-1)}\otimes\hat A\otimes\mathbb I^{\otimes(R-a)}$. Consequently, the clean Hamiltonian in the replicated Hilbert space is $\hat H_0^{(R)}=\sum_{a=1}^R\hat H_0^{(a)}$, and the disorder Liouvillian acting in replica $a$ is defined by $\mathcal L_i^{d,(a)}(t)X=[\hat n_i^{(a)}(t),X]$, where $\hat n_i^{(a)}=\hat c_i^{\dagger(a)}\hat c_i^{(a)}$ and $X$ denotes an operator in $\mathcal H^{\otimes R}$.
We define the replicated state as the tensor-product state $|\Psi^R_t\rangle\equiv|\psi_t\rangle^{\otimes R}\equiv \bigotimes_{a=1}^R|\psi_t\rangle^{(a)}$. We write the single-replica stochastic increment in the compact form $d|\psi_t\rangle=(\hat K_tdt+\sum_i\hat M_{i,t}dW_i^t)|\psi_t\rangle$, where $\hat K_t$ contains all deterministic $O(dt)$ contributions, including the Hamiltonian evolution, static disorder, and the It\^{o} drift, while $\hat M_{i,t}$ denotes the operator coupled to $dW_i^t$. For the present density-monitoring protocol, $\hat M_{i,t}=\hat n_i-\langle\hat n_i\rangle_t$. Applying the It\^{o} product rule to the replicated state gives
\begin{equation}
d|\Psi^R_t\rangle=\left[\sum_{a=1}^R\hat K_t^{(a)}dt+\sum_{a=1}^R\sum_i\hat M_{i,t}^{(a)}dW_i^t+\gamma dt\sum_{1\leq a<b\leq R}\sum_{i}\hat M_{i,t}^{(a)}\hat M_{i,t}^{(b)}\right]|\Psi^R_t\rangle.
\end{equation}
The second-order terms must be retained because different replicas are driven by the same Wiener increments, so that their cross products satisfy $dW_i^tdW_j^t=\gamma\delta_{ij}dt$. The $R$-replicated trajectory density matrix is defined as $\rho_t^{\otimes R}=|\Psi_t^R\rangle\langle\Psi_t^R|$, and applying the It\^{o} product rule once more, we obtain
\begin{equation}
\begin{aligned}
d\rho_t^{\otimes R}={}&\Bigg\{\sum_{a=1}^R[\hat K_t^{(a)}\rho_t^{\otimes R}+\rho_t^{\otimes R}\hat K_t^{(a)\dagger}]
+\gamma\sum_i\sum_{1\leq a<b\leq R}[\hat M_{i,t}^{(a)}\hat M_{i,t}^{(b)}\rho_t^{\otimes R}+\rho_t^{\otimes R}\hat M_{i,t}^{(b)\dagger}\hat M_{i,t}^{(a)\dagger}]
+\gamma\sum_i\sum_{a,b=1}^R\hat M_{i,t}^{(a)}\rho_t^{\otimes R}\hat M_{i,t}^{(b)\dagger}\Bigg\}dt\\
&+\sum_i\sum_{a=1}^R[\hat M_{i,t}^{(a)}\rho_t^{\otimes R}+\rho_t^{\otimes R}\hat M_{i,t}^{(a)\dagger}]dW_i^t.
\end{aligned}
\end{equation}
We then define the averaged replicated density matrix as $\rho_t^R\equiv\overline{\rho_t^{\otimes R}}$. The terms in the second line generate ket--ket and bra--bra couplings between different replicas, whereas the third line generates ket--bra couplings. In the Keldysh representation, these correspond respectively to same-branch and opposite-branch inter-replica couplings. The last term $\sum_{i=1}^L\sum_{a=1}^R[\hat M_{i,t}^{(a)}\rho_t^{\otimes R}+\rho_t^{\otimes R}\hat M_{i,t}^{(a)\dagger}]dW_i^t$ vanishes after we average over trajectories. In the static-disorder average, the disorder contribution to the Lindbladian is replaced by its replica sum, by $\mathcal L_i^d(t)\rho=[(\hat n_i)^I(t),\rho]\to \sum_{a=1}^R\mathcal L_i^{d,(a)}(t)\rho^R=\sum_{a=1}^R[(\hat n_i)^{I,{(a)}}(t),\rho^R]$, $\int_0^Tdt\int_0^Tdt'\mathcal L_i^d(t)\mathcal L_j^d(t')\to \sum_{a,b=1}^R\int_0^Tdt\int_0^Tdt'\mathcal L_i^{d,(a)}(t)\mathcal L_j^{d,(b)}(t')$.
The change in the averaged replicated density matrix $\rho^R_{t+dt}=\rho^R_t + d \rho^R_t $ over a time interval $dt$ is,
\begin{equation}\label{eq:rhoR}
\begin{aligned}
d\rho_t^R=dt\Bigg\{&
-i\sum_{a=1}^R[\hat H_0^{(a)},\rho_t^R]
-w\sum_{i=1}^L\sum_{a,b=1}^R\int_0^t dt'\,
[\hat n_i^{(a)}(t),[\hat n_i^{(b)}(t'),\rho_t^R]]
\\
&-\frac{\gamma}{2}\sum_{i=1}^L\sum_{a=1}^R
[\hat n_i^{(a)}(t),[\hat n_i^{(a)}(t),\rho_t^R]]
+\gamma\sum_{i=1}^L\sum_{1\leq a<b\leq R}
\{\hat{M}_i^{(a)}(t),\{\hat{M}_i^{(b)}(t),\rho_t^R\}\}
\Bigg\}.
\end{aligned}
\end{equation}
where $w=\frac{W^2}{3}$, and we have used $[\hat M_i^{(a)}(t),[\hat M_i^{(b)}(t'),\rho_t^R]]=[\hat n_i^{(a)}(t),[\hat n_i^{(b)}(t'),\rho_t^R]]$ for commutators. We introduce the grassmann field $\vec{\psi}=(\psi^{(1)},\psi^{(2)},\cdots \psi^{(R)})$, $\vec{\bar{\psi}}=(\bar{\psi}^{(1)},\bar{\psi}^{(2)},\cdots \bar{\psi}^{(R)})^T$. We adopt the symmetrized (Weyl) order to avoid the potential ambiguity arising from the discontinuity of $t=0^+$ and $t=0^-$ for the following Green function computations on Keldysh contour,  
$\hat{n}_i^{\pm,(a)}\to \frac{1}{2}+\bar{\psi}_i^{\pm,(a)}\psi_i^{\pm,(a)}$.
To represent the master equation Eq.~(\ref{eq:rhoR}) in terms of Keldysh path integral, we work on the Keldysh$\times$replica $2R\times 2R$ space . We map the bra and ket to Keldysh $-$ and $+$ contour, respectively, and define $u_i^{(a)}\equiv\bar{\psi}_i^{+,(a)}\psi_i^{+,(a)}$, $v_i^{(a)}\equiv\bar{\psi}_i^{-,(a)}\psi_i^{-,(a)}$. For the disorder terms, $[\hat n_i^{(a)}(t),[\hat n_i^{(b)}(t'),\rho_t^R]]\to (u^{(a)}_i(t)-v^{(a)}_i(t))(u^{(b)}_i(t')-v^{(b)}_i(t'))$. For the monitoring terms, $[\hat n_i^{(a)}(t),[\hat n_i^{(a)}(t),\rho_t^R]]\to \frac12-2u^{(a)}(t)v^{(a)}(t)$, where we have used $\hat{n}_i^2(t)=\hat{n}_i(t)$, and $\{\hat{M}_i^{(a)}(t),\{\hat{M}_i^{(b)}(t),\rho_t^R\}\}\to (u^{(a)}_i(t) + v^{(a)}_i(t)+1-2\langle \hat{n}_i\rangle_t)(u^{(b)}_i(t) + v^{(b)}_i(t)+1-2\langle \hat{n}_i\rangle_t)$.

Next, we combine fields in $R$-Replica $\times$ Keldysh space, 
$\Psi=(\vec{\psi}^+,\vec{\psi}^-)^T$, $\bar{\Psi}=(\vec{\bar{\psi}}^+,\vec{\bar{\psi}}^-)$,
where both fields $\Psi$ and $\bar{\Psi}$ contain $2R$ elements. We use the property  $\sum_a (u^{(a)}+v^{(a)})=-\Tr(\Psi\bar{\Psi}), \sum_a (u^{(a)}-v^{(a)})=-\Tr(\sigma^z_K\Psi\bar{\Psi})$, where the index "K" in $\sigma_K^z$ means the Pauli matrix acts only in the $2\times 2$ Keldysh space. As a result, $\sum_{a,b}(u^{(a)}+v^{(a)})(u^{(b)}+v^{(b)})=\Tr((\Psi\bar{\Psi})_{aa}(\Psi\bar{\Psi})_{bb})=-
\Tr((\Psi\bar{\Psi})_{ab}(\Psi\bar{\Psi})_{ba})=-\Tr((\Psi\bar{\Psi})^2)$, similarly, we have, $\sum_{a,b}(u^{(a)}-v^{(a)})(u^{(b)}-v^{(b)})=-\Tr((\sigma^z_K\Psi\bar{\Psi})^2)$. The full action reads as,
\begin{equation}\label{eq:action}
    iS[\bar{\Psi},\Psi]= \Tr\left(i \bar{\Psi}\sigma^z_K(i\partial_t -H_0(x))\Psi -\frac{\gamma}{2} \delta(t-t') (\Psi\bar{\Psi})^2+\frac{w}{2} (\sigma^z_K\Psi\bar{\Psi})^2 -(R-1)\gamma\delta(t-t')(1-2\langle \hat{n}\rangle_t)(\Psi\bar{\Psi})\right)
\end{equation}
where $\Tr(\cdots)$ denotes the trace over the replica$\times$Keldysh space, the internal-space index $i=1,2,\ldots,L$, and the bilocal time variables $(t,t')$. The squares in $(\sigma^z_K\Psi\bar{\Psi})^2$ and $(\Psi\bar{\Psi})^2$ denote matrix squares. The quantity $\langle\hat n\rangle_t$ is understood as an operator diagonal in real space, whose $i$th diagonal element is $\langle\hat n_i\rangle_t$; the summation over $i$ is therefore included in $\Tr$. In particular, the tadpole structure $\Tr[\delta(t-t')(\Psi_i\bar{\Psi}_i)]$ simply reduces to the single-time integral $\Tr_{E}\sum_i\int dt(\Psi_i\bar{\Psi}_i)(t)$, where $\Tr_{E}$ denotes the trace over the external replica and Keldysh indices only. We have discarded all constant terms in the action.

 Next, we introduce the auxiliary field $\mathcal{G}=i \Psi\bar{\Psi}$ by the delta function $1=\int D \mathcal{G}\delta(\mathcal{G}-i \Psi\bar{\Psi})$, and the auxiliary field $Q$ is introduced in the exponential form of delta function,
\begin{equation}\label{eq:HS}
1=\int D\mathcal{G} DQ \exp\Big( i\Tr \left(Q(\mathcal{G}-i\Psi\bar{\Psi})\right) \Big)
=\int D\mathcal{G}  DQ\exp \Big(i(\gamma+w)\Tr\left(\mathcal{G}Q+i\bar{\Psi} Q\Psi \right)\Big)
\end{equation}
where we have used the fact $\Tr(Q\Psi\bar{\Psi})=-\Tr(\bar{\Psi} Q\Psi)$, since $Q$ is a bosonic collective field and $Q_{ij}\Psi_j \bar{\Psi}_i=-\bar{\Psi}_i Q_{ij}\Psi_j$. Either sign of the $\pm i$ prefactor may be used; it only changes the sign convention for the saddle, $Q_{\mathrm{saddle}}=\pm\sigma_K^z$. We adopt the positive-$i$ convention. We rescale $Q$ by $(\gamma+w)$ and absorb the field-independent Jacobian into the normalization of the functional measure. 

Physically, the collective slow modes carry a small external momentum $q$ and frequency $\Omega$. Consider the four-fermion vertex $\Tr[(\Psi\bar{\Psi})^2]$, with the fermionic fields ordered as $\Psi_a(k_1)\bar{\Psi}_b(k_2)\Psi_b(k_3)\bar{\Psi}_a(k_4)$ and $k_1-k_2+k_3-k_4=0$. We denote by $\mathcal G_d$ and $\mathcal G_x$ the projections of the same collective field $\mathcal G$ onto the direct and exchanged slow-momentum routings, respectively. The direct pairing is characterized by the small transfer $q_d=k_1-k_2=k_4-k_3$ and gives the matrix channel $-\Tr(\mathcal G_d^2)$, whereas the exchanged pairing is characterized by $q_x=k_1-k_4=k_2-k_3$ and gives the trace channel $[\Tr(\mathcal G_x)]^2$. The frequency transfers are defined analogously. In either pairing, the corresponding fermionic center momenta remain unrestricted and are integrated over the full spectrum. The two slow sectors therefore occupy distinct leading regions of the four-leg momentum space. Their overlap requires both $q_d$ and $q_x$ to be small and is neglected as a subleading phase-space contribution in the leading slow-mode approximation. Before this projection, the Grassmann identity $\Tr[(\Psi\bar{\Psi})^2]=-\Tr(\mathcal G^2)=[\Tr(\mathcal G)]^2$ makes the two expressions alternative representations of the same full vertex. After the projection, however, the Grassmann rearrangement exchanges the momentum transfer constrained to be small and therefore does not commute with the slow-mode projection. Using either representation alone would retain only one of the two slow sectors. We therefore use $\left.\Tr[(\Psi\bar{\Psi})^2]\right|_{\mathrm{slow}}\simeq-\Tr(\mathcal G_d^2)+[\Tr(\mathcal G_x)]^2$, where the neglected correction comes from the overlap between the two projected slow sectors.

It can be shown that the two trace structures, $\Tr\mathcal G$ and $\Tr(\mathcal G\sigma_K^z)$, are orthogonal linear combinations of the replica traces of the two Keldysh-diagonal components of $\mathcal G$. The replica trace eliminates all traceless replica components, while the insertion of $\sigma_K^z$ changes only the relative sign between the two Keldysh-diagonal components, so that the two trace structures select their sum and difference, respectively. Their corresponding quadratic channels therefore lie entirely within the $U(1)\times U(1)$ replica-symmetric sector, which is precisely the $R=1$ reduction of the full $U(R)\times U(R)$ symmetry established below, and introduce no additional replicon channel. 
We will show below that the saddle selected by our analysis is replica symmetric at $R=1$ and that the nontrivial long-time contribution to the steady-state entanglement entropy is governed solely by the leading-order replicon fluctuations around this saddle. Consequently, the replica-singlet channels built from $\Tr\mathcal G$ and $\Tr(\mathcal G\sigma_K^z)$ introduce no additional long-time degrees of freedom in the replicon sector. At leading order, we therefore make the approximation
\begin{equation}
\Tr(\mathcal{G}_{\mathrm{slow}}^2)\sim-\big[\Tr(\mathcal{G}_{\mathrm{slow}})\big]^2,\quad
\Tr\big[(\mathcal{G}_{\mathrm{slow}}\sigma_K^z)^2\big]\sim-\big[\Tr(\mathcal{G}_{\mathrm{slow}}\sigma_K^z)\big]^2,
\end{equation}
where $\mathcal G_{\mathrm{slow}}$ denotes the projection of $\mathcal G$ onto the slow-mode sector; within the leading-order approximation, we do not distinguish between the direct and exchange sectors.
Within this approximation, the coefficients of $\Tr(\mathcal G^2)$ and $\Tr[(\mathcal G\sigma_K^z)^2]$ are each doubled relative to the full-spectrum calculation. Hereafter, $\mathcal G$ and $Q$ are understood to denote their slow-mode-projected components, and we omit the subscript ``slow'' for notational simplicity. We then derive the slow-mode-projected collective-field action $iS[\mathcal G,Q]$ from the fermionic action $iS[\bar\Psi,\Psi]$,
\begin{equation}
\label{eq:SGQ}
 iS[\mathcal{G},Q]=\Tr\log\Big(i\sigma^z_K(i\partial_t -H_0(x))-(\gamma+w)Q\Big) +\Tr\Big(i(\gamma+w)\mathcal{G}Q+\gamma\delta(t-t') \mathcal{G}^2-w(\mathcal{G}\sigma^z_K)^2+(R-1)\gamma i\delta(t-t')(1-2\langle \hat{n}\rangle_t)\mathcal{G}\Big)
\end{equation}

Before performing the Gaussian integration over $\mathcal G$, we separate the quadratic auxiliary-field terms into their equal-time and non-equal-time sectors in the bilocal time-kernel space,
$\mathcal G=\mathcal G^{t=t'}+\mathcal G^{t\neq t'}$ and
$Q=Q^{t=t'}+Q^{t\neq t'}$.
The two sectors have disjoint support and are therefore orthogonal under the full kernel trace. For example,
$\Tr\left[\delta(t-t')\mathcal G^2\right]\equiv\Tr\left[\delta(t-t')\mathcal G(t,t')\mathcal G(t',t)\right]$
is supported only at equal times and reduces to
$\int dt \Tr_E\left[\mathcal G(t,t)^2\right]$. Assuming time-translation invariance, the equal-time component is denoted by $\mathcal G(0)$, whereas the non-equal-time component takes the form
$\mathcal G(\Delta t)\mathcal G(-\Delta t)$ with $\Delta t=t-t'\neq0$.
Consequently, the disorder-induced interaction contributes to both the equal-time and non-equal-time sectors, whereas the monitoring-induced interaction contributes only to the equal-time sector.
We then decompose the interacting part in Eq.~(\ref{eq:SGQ}) into the $t=t'$ and $t\neq t'$ channel, while the term $\Tr\log(\cdots)$ is kept intact at this stage,
\begin{equation}\label{eq:actionS}
\begin{aligned}
iS^{\mathrm{int}}_{t=t'}[\mathcal{G}^{t=t'},Q^{t=t'}]&=\Tr\Big(i(\gamma+w)\mathcal{G}^{t=t'}Q^{t=t'}+\gamma(\mathcal{G}^{t=t'})^2-w(\mathcal{G}^{t=t'}\sigma^z_K)^2+i(R-1)\gamma(1-2\langle\hat n\rangle_t)\mathcal{G}^{t=t'}\Big),\\
iS^{\mathrm{int}}_{t\neq t'}[\mathcal{G}^{t\neq t'},Q^{t\neq t'}]&=\Tr\Big(i(\gamma+w)\mathcal{G}^{t\neq t'}Q^{t\neq t'}-w(\mathcal{G}^{t\neq t'}\sigma^z_K)^2\Big).
\end{aligned}
\end{equation}
Here, the superscript $\mathrm{int}$ indicates that only the auxiliary-field terms generated by the quartic interaction, together with the tadpole term, are included. The tadpole $\Tr[\delta(t-t')\mathcal{G}(t,t')]$ selects only the equal-time $t=t'$ sector of $\mathcal{G}$.

The main purpose of the derivation in this section is to obtain the saddle-point equation and determine its physical saddle configuration. A physically plausible saddle is supported in the equal-time sector, $t=t'$. Hereafter, unless explicitly stated otherwise, $\mathcal G$ and $Q$ without temporal labels denote $\mathcal G^{t=t'}$ and $Q^{t=t'}$, respectively. The fermionic $\Tr\log(\cdots)$ term is a single smooth functional of the full bilocal field $Q(t,t')$. Within the leading-order saddle-point approximation, we can therefore evaluate this term at the equal-time saddle and smoothly extend the fluctuation expansion around that saddle to the non-equal-time sector. In contrast, because the interaction part contains both the equal-time monitoring contribution and the bilocal temporal coupling induced by disorder, it changes discontinuously between the equal-time and non-equal-time sectors. We will show below that this difference generates a positive bare mass for the non-equal-time component within the leading saddle-dominance approximation, allowing it to be integrated out without altering the infrared RG flow of the remaining slow-mode theory, apart from finite local corrections to its initial couplings. Accordingly, we now focus on $S^{\mathrm{int}}_{t=t'}$ when deriving the saddle configuration.

Then we decompose the Keldysh$\times$Replica space into two orthogonal channels $\mathcal G_\pm$ and $Q_\pm$,
\begin{equation}\label{eq:SplitKR1}
\mathcal G_\pm=\frac{1}{2}\left(\mathcal G\pm\sigma_K^z\mathcal G\sigma_K^z\right),
\qquad
Q_\pm=\frac{1}{2}\left(Q\pm\sigma_K^zQ\sigma_K^z\right).
\end{equation}
Here, $\mathcal G_+$ and $Q_+$ are block diagonal in Keldysh space, whereas $\mathcal G_-$ and $Q_-$ are block off-diagonal. Defining $\mathcal G_z\equiv\sigma_K^z\mathcal G$ and $Q_z\equiv\sigma_K^zQ$, we have
\begin{equation}\label{eq:SplitKR2}
\begin{aligned}
&\mathcal G=\mathcal G_++\mathcal G_-,
\quad Q=Q_++Q_-,
\quad \sigma_K^z\mathcal G_\pm\sigma_K^z=\pm\mathcal G_\pm,
\quad \sigma_K^zQ_\pm\sigma_K^z=\pm Q_\pm,\\
&\Tr(\mathcal G^2)=\Tr(\mathcal G_+^2+\mathcal G_-^2),
\quad \Tr(\mathcal G_z^2)=\Tr(\mathcal G_+^2-\mathcal G_-^2),
\quad \Tr(Q^2)=\Tr(Q_+^2+Q_-^2),
\quad \Tr(Q_z^2)=\Tr(Q_+^2-Q_-^2).
\end{aligned}
\end{equation}
Since $Q_+\mathcal G_-$ and $Q_-\mathcal G_+$ are block off-diagonal in Keldysh space, their traces vanish. Consequently, $\Tr(Q\mathcal G)=\Tr(Q_+\mathcal G_+)+\Tr(Q_-\mathcal G_-)$, and the interaction action separates as,
\begin{equation}\label{eq:SeparateChannel}
\begin{aligned}
&iS^{t=t',\mathrm{int}}[Q,\mathcal G]=iS^{t=t',\mathrm{int}}_+[Q_+,\mathcal G_+]+iS^{t=t',\mathrm{int}}_-[Q_-,\mathcal G_-]\\
&iS^{t=t',\mathrm{int}}_+[Q_+,\mathcal G_+]
=\Tr\left[(\gamma-w)\mathcal G_+^2+i(\gamma+w)Q_+\mathcal G_+
+i(R-1)\gamma(1-2\langle\hat n\rangle_t)\mathcal G_+\right],\\
&iS^{t=t',\mathrm{int}}_-[Q_-,\mathcal G_-]
=\Tr\left[(\gamma+w)\mathcal G_-^2+i(\gamma+w)Q_-\mathcal G_-\right].
\end{aligned}
\end{equation}
In the $t=t'$ sector, the tadpole couples only to the $+$ channel because $1-2\langle\hat n\rangle_t$ is proportional to the identity in Keldysh space and $\Tr[(1-2\langle\hat n\rangle_t)\mathcal G_-]=0$. We assume $\gamma>w$ from now on, so that all Gaussian coefficients in both $\pm$ channel of the equal-time sector are positive, hence the corresponding Gaussian integrals do not converge on the original Hermitian contours. The divergence requires the analytical continuation of the matrix field $\mathcal{G}$ from the real axis to the imaginary axis, which can be done as follows. Taking the $-$ channel as an example, we first extract the common $(\gamma+w)$ factor and expand the independent field components as
$\mathcal G=\sum_\mu g_\mu T_\mu$ and $Q=\sum_\mu q_\mu T_\mu$, where $T_\mu$ form an orthonormal basis of the $-$ channel in replica$\times$Keldysh space, $\Tr(T_\mu T_\nu)=\delta_{\mu\nu}$ , while all remaining space-time
indices are left implicit in $\Tr$. The quadratic kernel of the $\mathcal G$ integral is therefore
$K_{\mu\nu}=2\delta_{\mu\nu}$.
For each independent mode, we define the Gaussian integral by analytically continuing the quadratic kernel along the path parametrized by $\phi$,
\begin{equation}
K_{\mu\nu}(\phi)= 2e^{i\phi}\delta_{\mu\nu},
\qquad
g_\mu=e^{i(\pi-\phi)/2}x_\mu,
\qquad
\phi:\pi\rightarrow0,
\end{equation}
while keeping the linear source
$q_\mu$ fixed. Along this continuation,$
\frac{1}{2}\sum_{\mu,\nu}g_\mu K_{\mu\nu}(\phi)g_\nu=-\sum_\mu x_\mu^2$, 
so that the convergent real-axis Gaussian at $\phi=\pi$ is continuously continued to the imaginary-axis contour
$\mathcal G=iX$ at $\phi=0$. Since the integrand is entire in every independent component $g_\mu$, no finite pole or branch cut is crossed during this continuation.
The auxiliary-field insertion is understood in the finite-$\varepsilon$ regularization introduced in Appendix B of Ref.~\cite{poboiko2023}, which fixes the original Fourier representation and normalization of the delta functional. The analytic continuation described above is a separate convergence prescription for the otherwise divergent $\mathcal G$ Gaussian integral. It leaves the linear constraint term unchanged and is performed before taking the limit $\varepsilon\to0^+$.

We then integrate $\mathcal{G}$ along the imaginary axis in both $\pm$ channels of the $t=t'$ sector and along the real axis in the $t\neq t'$ sector,
\begin{equation}
\begin{aligned}\label{eq:FreqAction}
&iS^{t\neq t',\mathrm{int}}[Q]=-\frac{(\gamma+w)^2}{4w}\Tr\left[\left(Q^{t\neq t'}\sigma^z_K\right)^2\right]\\
&iS^{t=t',\mathrm{int}}[Q_+,Q_-] =\Tr\left[\frac{(\gamma+w)^2}{4(\gamma-w)}
\left(Q_+ + \frac{(R-1)\gamma}{\gamma+w}(1-2\langle\hat n\rangle_t)\right)^2
+\frac{\gamma+w}{4}Q_-^2\right].
\end{aligned}
\end{equation}
which gives the interaction part of the effective action for $Q$.

We next consider the equal-time $t=t'$ sector and take the replica symmetric sector $R=1$ to select the saddle configuration. This restriction is used only to determine the physical saddle configuration, and the replica-dependent saddle manifold and its fluctuations are analyzed separately below. 
Rewriting the quadratic terms for the $iS^{t=t',\mathrm{int}}[Q]$ term in the $Q,Q_z$ basis using Eq.~(\ref{eq:SplitKR2}), we obtain
\begin{equation}\label{eq:saddleAction1}
iS_{\mathrm{saddle}}[Q]
=\Tr\log\left[i\sigma_K^z(i\partial_t-H_0(x))-(\gamma+w)Q\right]
+\Tr\left[\frac{\gamma(\gamma+w)}{4(\gamma-w)}Q^2+\frac{w(\gamma+w)}{4(\gamma-w)}(\sigma_K^zQ)^2\right].
\end{equation}
We now impose the causality condition by performing the Keldysh rotation in Keldysh space, using the convention
$\Psi_{\pm}=\tilde{U}\Psi_K$ and $\bar{\Psi}_{\pm}=\bar{\Psi}_K \tilde{U}\sigma_K^z$, with
$\tilde{U}=\frac{1}{\sqrt{2}}\begin{pmatrix}1&1\\1&-1\end{pmatrix}$.
Accordingly, a bilinear kernel transforms as $A_K=\tilde{U}\sigma_K^zA_{\pm}\tilde{U}$, so that
$\mathbf{1}_K\to\sigma_K^x$ and $\sigma_K^z\to\mathbf{1}_K$.
We choose the infinitesimal prescription such that the rotated inverse propagator contains the standard causal regulator $i0\sigma_K^z$. We then define the dressed Green function by
\begin{equation}
G^{-1}[Q]=i\partial_t-H_0(x)+i0\sigma_K^z+i(\gamma+w)Q \qquad G=\begin{pmatrix}G^R&G^K\\0&G^A\end{pmatrix}_K
\end{equation}
The Keldysh rotation brings the dressed Green function into upper-triangular form. For notational simplicity, we henceforth suppress the subscript $K$ on the rotated field $Q_K=\tilde{U}\sigma_K^zQ\tilde{U}$. The Keldysh-rotated action then reads,
\begin{equation} \label{eq:saddleAction2}
iS_{\mathrm{saddle}}^K[Q]=\Tr\log\left(G^{-1}[Q]\right)+\Tr\left[\frac{\gamma(\gamma+w)}{4(\gamma-w)}(Q\sigma_K^x)^2+\frac{w(\gamma+w)}{4(\gamma-w)}Q^2\right]
\end{equation}
where the constant $\Tr\log(i)$ has been absorbed into the normalization of the functional measure.

For generic replica number $R$, the term $\Tr[(Q\sigma_K^x)^2]$ mixes the retarded and advanced causal branches in the saddle-point equation. Consequently, the usual branchwise solution in terms of $G^R$ and $G^A$ does not close, and a simple causal saddle cannot be obtained in this form. In the replica limit $R\to1$, however, the replica-symmetric saddle reduces to a $2\times2$ Keldysh problem, for which the exact identity
$\Tr_K[(Q\sigma_K^x)^2]+\Tr_K(Q^2)=[\Tr_K(Q\sigma_K^x)]^2+[\Tr_K(Q)]^2$
holds. The causality constraint requires $\Tr_K(Q)=0$ within the saddle sector.

Continuous monitoring heats the energy distribution to infinite temperature, while the conserved filling $n$ remains fixed, so that the distribution function becomes energy independent, $F(\varepsilon)=1-2n$. We choose the homogeneous half-filled saddle, $n=1/2$, for which $F=0$ and $\Tr_K(Q\sigma_K^x)=0$. Within this restricted saddle sector, we therefore have
$\Tr[(Q\sigma_K^x)^2]=-\Tr(Q^2)$, and the action is further simplified to,
\begin{equation}\label{eq:saddleAction3}
iS_{\mathrm{saddle}}^K[Q]=\Tr\log\left(i\partial_t-H_0(x)+i0\sigma_K^z+i(\gamma+w)Q\right)-\Tr\left(\frac{\gamma+w}{4}Q^2\right),
\end{equation}
where we have restricted the action to the homogeneous half-filled $R=1$ saddle sector specified above. No corresponding simplification is assumed for general replica-symmetric Keldysh fluctuations or away from half filling.
Transforming the spatially homogeneous and time-local saddle equation to momentum-frequency space, we obtain
\begin{equation}\label{eq:SCBA}
\begin{aligned}
Q_s^{R/A}
&=2i\int_{-\pi}^{\pi}\frac{dk}{2\pi}\int_{-\infty}^{\infty}\frac{d\omega}{2\pi}
\left[\omega-\xi_k+i(\gamma+w)Q_s^{R/A}\pm i0\right]^{-1}
=\mathrm{sgn}(Q_s^{R/A}),\\
Q_s^R&=1,\qquad Q_s^A=-1,\qquad Q_s=\sigma_K^z.
\end{aligned}
\end{equation}
Here, $\omega$ is the frequency conjugate to the relative time, while $\xi_k$ is the single-particle dispersion, i.e., the eigenvalue of $H_0$ in momentum space. The energy integral is evaluated using the causal equal-time prescription inherited from the retarded and advanced Green functions.
\section{Symmetry classification and the mean-field diffusive dynamics at the saddle}\label{app:NLSM2}
We first trace the symmetry structure from the Hamiltonian to the action and finally to the saddle manifold.

The free Hamiltonian $H_0$ has two symmetries: \par
(1). Time-Reversal (TRS): $H_0=H_0^*$;\par
(2). Sublattice (CS): $\{H_0,\hat{S}\}=0$, $\hat{S}=\mathbf{1}_{L/2}\otimes \sigma^z$. The even integer $L$ is the system size. \par
The non-zero disorder strength $w\neq 0$ explicitly breaks the CS symmetry, while the non-zero monitoring strength $\gamma\neq 0$, in terms of the real diffusive terms in Eq.~(\ref{eq:SSE}), breaks TRS explicitly. Consequently, the monitored disordered system has no additional symmetry, whereas in the clean monitored system the remaining CS is lifted to the PHS$^\dagger$ constraint of the doubled Lagrangian operator, $\hat S\hat L^\dagger\hat S^{-1}=-\hat L$.

We next classify the quadratic, unaveraged replicated Keldysh action $iS_{\mathrm{unaveraged}}[\bar{\Psi},\Psi]$ for a fixed trajectory and a fixed disorder realization. For this purpose, we formally define $iS_{\mathrm{unaveraged}}$ by replicating the fixed-trajectory stochastic Schr\"odinger equation in Eq.~(\ref{eq:SSE}), using the same fixed stochastic realization in all replicas, lifting its bra--ket structure to the Keldysh action, and discarding the deterministic interaction terms generated by products of the It\^o increments $dW_i^t dW_j^t\propto \delta_{ij}dt$. This construction is equivalent, for symmetry-classification purposes, to the classification of the doubled Lagrangian operator in Ref.~\cite{poboiko2024}.

The action is bilinear in $\bar{\Psi}$ and $\Psi$ and contains the two channels $\sigma_K^z\Psi\bar{\Psi}$ and $\Psi\bar{\Psi}$ in replica$\times$Keldysh space. We consider a global transformation $\Psi\to \hat{U}\Psi$, $\bar{\Psi}\to\bar{\Psi}\hat{U}^\dagger$, with $\hat{U}\in U(2R)$. The subgroup that leaves the reference quadratic action invariant is $H=\{h\in U(2R)\mid h\sigma_K^zh^\dagger=\sigma_K^z\}$. Together with $hh^\dagger=\mathbf I_{2R}$, this condition implies $[h,\sigma_K^z]=0$, so that $h$ contains only two block-diagonal $R\times R$ components and $H\simeq U(R)\times U(R)$. Since the transformed quadratic action $iS_{\mathrm{unaveraged}}[\hat{U}\Psi,\bar{\Psi}\hat{U}^\dagger]$ is unchanged under the right equivalence $\hat{U}\sim \hat{U}h$, $h\in H$, the inequivalent quadratic forms are parametrized by $U(2R)/(U(R)\times U(R))$, corresponding to Cartan class AIII. In the clean case, the additional PHS$^\dagger$ symmetry imposes a reality condition, reducing $U(2R)$ to $O(2R)$ and $H$ to $O(R)\times O(R)$. The corresponding classification space is therefore $O(2R)/(O(R)\times O(R))$, namely Cartan class BDI, as worked out in Ref.~\cite{poboiko2024}. These AIII and BDI classifications refer to the fixed-trajectory quadratic action.

In the main text, at leading order, we characterize the entanglement entropy $S_A=-\Tr_A(\rho_A\log\rho_A)$ through the second cumulant of the particle number in a subregion $A$,
$$
C_A^{(2)}=\left\langle(\delta N_A)^2\right\rangle=\sum_{x,y\in A}\left\langle\delta n(x)\delta n(y)\right\rangle,\qquad N_A=\sum_{x\in A}n(x).
$$
For noncoincident points $x\neq y$, the Weyl-ordering shifts contribute only local contact terms, and the connected density correlation is mapped to the bilocal collective-field correlation, $\left\langle\delta n(x)\delta n(y)\right\rangle_c\to(\sigma_K^z)_{\alpha\beta}(\sigma_K^z)_{\gamma\delta}\left\langle\delta\mathcal G_{\beta\gamma}(x-y)\delta\mathcal G_{\delta\alpha}(y-x)\right\rangle=\left\langle\Tr_{E}\!\left[\sigma_K^z\delta\mathcal G(x-y)\sigma_K^z\delta\mathcal G(y-x)\right]\right\rangle$, where $\alpha,\beta,\gamma,\delta$ are combined replica$\times$Keldysh indices. By varying the action $iS[\mathcal G,Q]$ in Eq.~(\ref{eq:SGQ}), the saddle-point equations express $\delta\mathcal G$ as a linear combination of $\delta Q$ and $\delta(\sigma_K^zQ\sigma_K^z)$, using the standard $+/-$ basis before the Keldysh rotation. Therefore, it is sufficient to consider the $\delta Q$ sector.

Taking the disordered case as an example, we define $Q=T\Lambda T^{-1}$, with $T=e^\mathcal{W}\in U(R)\times U(R)$, and obtain the linear fluctuation $\delta Q=[\mathcal{W},\Lambda]+O(\mathcal{W}^2)$. Generators satisfying $[\mathcal{W},\Lambda]=0$ belong to the stabilizer and do not generate density fluctuations. It follows directly that the stabilizer is the diagonal group $U(R)_{\mathrm{diag}}\equiv\{(U,U)\mid U\in U(R)\}$, which is isomorphic to $U(R)$, so that the saddle manifold is $\mathcal M=(U(R)\times U(R))/U(R)\simeq U(R)$. For the clean case, the saddle manifold can be obtained by the same stabilizer-group method, giving $\mathcal M\simeq U(2R)/Sp(2R)$ \cite{poboiko2024}. We note that the correspondence between the global symmetry group of the replicated action and the saddle-point manifold is well established in the theory of Anderson localization \cite{evers08}. From now on, we focus on the disordered case $w\neq0$.

We now decompose the theory into the replica-symmetric and replicon, or replica-off-diagonal, sectors. In the replica-symmetric sector associated with the global $U(1)$ symmetry, the calculation of $C(r)$ yields diffusive dynamics governed by the kernel $\mathcal Dq^2-i\Omega$. Here, $q$ and $\Omega$ denote the external momentum and frequency, respectively, and $\mathcal D$ is the diffusion constant. Consequently, the contribution of this replica-symmetric diffusive sector to the entanglement entropy vanishes in the long-time regime. The derivation proceeds similarly to Ref.~\cite{poboiko2023}. For the replicon sector, we use the Lie-algebra decomposition $\mathfrak u(R)=\mathfrak u(1)\oplus\mathfrak{su}(R)$ and retain the traceless field satisfying $i\Phi\in\mathfrak{su}(R)$. Since $Q$ has a $2R\times2R$ matrix structure in replica$\times$Keldysh space, we must also specify its direction within the $2\times2$ Keldysh block. From the action in Eq.~(\ref{eq:saddleAction2}), only $\mathbf 1_{2,K}$ and $\sigma_K^x$ are exact symmetry directions in the Keldysh-rotated basis, whereas the $\sigma_K^y$ and $\sigma_K^z$ directions lead to massive modes that affect the short-time dynamics. Of the two exact symmetry directions, $\mathbf 1_{2,K}$ commutes with the saddle $\Lambda$ and therefore does not contribute to saddle fluctuations. We thus parameterize the fluctuations along the $\sigma_K^x$ direction,
\begin{equation}\label{eq:Replicon}
Q=T\Lambda T^{-1},\qquad T=e^{i\Phi\sigma_K^x/2},\qquad i\Phi\in\mathfrak{su}(R),\qquad \delta Q=\frac{i}{2}\Phi\left[\sigma_K^x,\sigma_K^z\right]=\Phi\sigma_K^y.
\end{equation}
where $Q=T\Lambda T^{-1}$ parametrizes the Goldstone manifold around the saddle $\Lambda$, and $T$ characterizes the nontrivial directions associated with the spontaneously broken internal symmetry. Taking the temporal indices as an example, this parametrization allows continuous deformations away from the equal-time $t=t'$ saddle, $Q(t,t')=\int dt_1dt_2 T(t,t_1)\Lambda\delta(t_1-t_2)T^{-1}(t_2,t')$. The resulting non-equal-time components may nevertheless be lifted by the interaction terms, as shown below. Since $\Lambda^2=\mathbf 1_{2R}$, the trace over the replica$\times$Keldysh indices gives $\Tr_E(Q^2)=2R$, which is constant and can therefore be discarded.  We define the dressed Green function $G_\Lambda=\left(G_0^{-1}+i(\gamma+w)\Lambda\right)^{-1}$, where $G_0^{-1}=i\partial_t-H_0+i0\sigma_K^z$.
The trace-log term can then be rewritten as
\begin{equation}
\begin{aligned}
iS_{\mathrm{saddle}}^K[Q]
&=\Tr\log\left(G_0^{-1}+i(\gamma+w) T\Lambda T^{-1}\right)=\Tr\log\left(T^{-1}G_0^{-1}T+i(\gamma+w)\Lambda\right)\\
&=\Tr\log\left(G_\Lambda^{-1}+T^{-1}[G_0^{-1},T]\right)
=\Tr\log\left(\mathbf 1+G_\Lambda T^{-1}[G_0^{-1},T]\right).
\end{aligned}
\end{equation}
The last equality follows from the fermionic Keldysh causality identity $\Tr\log G_\Lambda=0$, since its variation generates only closed retarded-retarded and advanced-advanced contractions, both of which vanish.

To extract the spatial gradients, we write $H_0(x,x')=H_0(r)$ for a translationally invariant Hamiltonian, where $r=x-x'$, and introduce the center coordinate $X=(x+x')/2$. Treating $\partial_X$ as the small expansion parameter, the commutator kernel has the following leading slow-field expansion,
\begin{equation}
\begin{aligned}
[H_0,T](X,r)=H_0(r)\left[T\left(X-\frac r2\right)-T\left(X+\frac r2\right)\right]=-rH_0(r)\,\partial_XT(X)+O\!\left(r^3\partial_X^3T\right)=[H_0,\hat x](r)\,\partial_XT(X)+O\!\left(r^3\partial_X^3T\right).
\end{aligned}
\end{equation}
We henceforth relabel $X$ as $x$, define $a_\mu=T^{-1}\partial_\mu T$, and introduce the velocity operator $\hat v=i[H_0,\hat x]$. By the Heisenberg equation, $\hat v=d\hat x/dt$ is the single-particle velocity operator; in momentum space, $\hat v(k)=v_k=\partial_k\xi_k$. Retaining the leading spatial-gradient vertex, we obtain $T^{-1}[G_0^{-1},T]=i(a_t+\hat v a_x)+i0\,T^{-1}[\sigma_K^z,T]$. The infinitesimal regulator selects the causal saddle and is then taken to zero. The trace-log term relevant to the gradient expansion is therefore
\begin{equation}
iS_{\mathrm{saddle}}^K[Q]=\Tr\log\left[\mathbf 1+iG_\Lambda(a_t+\hat v a_x)\right].
\end{equation}

Expanding with $\Tr\log(\mathbf 1+X)=\Tr X-\frac12\Tr X^2+O(X^3)$, we obtain
\begin{equation}
\begin{aligned}\label{eq:grad}
iS_{\mathrm{saddle}}^K[Q]
={}&i\Tr(G_\Lambda a_t)+i\Tr(G_\Lambda\hat v a_x)
+\frac12\Tr(G_\Lambda a_tG_\Lambda a_t)
+\frac12\Tr(G_\Lambda\hat v a_xG_\Lambda\hat v a_x)\\
&+\frac12\Tr\left(G_\Lambda a_tG_\Lambda\hat v a_x
+G_\Lambda\hat v a_xG_\Lambda a_t\right)+O(\partial^3).
\end{aligned}
\end{equation}
Here, $O(\partial^3)$ denotes terms of total gradient order three and higher in the trace-log expansion. Throughout this calculation, $\Tr$ denotes the full trace over replica, Keldysh, and spacetime variables; when momentum and frequency integrations are written explicitly, it denotes the trace over the remaining variables.

Next, we introduce $U=e^{i\Phi}\in SU(R)$ and decompose $T$ using the projectors onto the $\sigma_K^x$ eigenspaces,
\begin{equation}
T=\hat P_+^xU^{1/2}+\hat P_-^xU^{-1/2},\qquad
\hat P_\pm^x=\frac{\mathbf 1_K\pm\sigma_K^x}{2},\qquad
a_\mu=\hat P_+^xU^{-1/2}\partial_\mu U^{1/2}
+\hat P_-^xU^{1/2}\partial_\mu U^{-1/2}.
\end{equation}
Using $\int_{k,\omega}G_\Lambda(k,\omega)=-i\Lambda/2$, where $\int_{k,\omega}\equiv\int\frac{dk}{2\pi}\frac{d\omega}{2\pi}$, and the fact that $\Lambda\hat P_\pm^x$ is traceless in Keldysh space, the first-order temporal term in Eq.~(\ref{eq:grad}) vanishes: $i\Tr(G_\Lambda a_t)=\frac12\Tr(\Lambda a_t)=0$.

For the first-order spatial term, the momentum-space integrand is proportional to $v_kG_\Lambda(k,\omega)a_x$. For nearest-neighbor hopping, $\xi_{-k}=\xi_k$, $v_{-k}=-v_k$, and $G_\Lambda(-k,\omega)=G_\Lambda(k,\omega)$; hence the momentum integral vanishes by parity. The mixed $a_ta_x$ terms contain one velocity insertion and vanish by the same parity argument in the slow-field limit.

For the remaining quadratic terms, we define the projectors onto the retarded and advanced sectors by $\widehat P_\pm=(\mathbf 1_{2R}\pm\Lambda)/2$ and decompose $G_\Lambda=\widehat P_+G_\Lambda^R+\widehat P_-G_\Lambda^A$. Closed $RR$ and $AA$ convolution traces vanish by causality, so only the $RA$ and $AR$ components remain. For the temporal quadratic term,
\begin{equation}
\begin{aligned}
\Tr(G_\Lambda a_tG_\Lambda a_t)
&=\int_{q,\Omega}\int_{k,\omega}\Big[
G_\Lambda^R(\omega,k)G_\Lambda^A(\omega+\Omega,k+q)
\Tr_E\left(\widehat P_+a_t(q,\Omega)\widehat P_-a_t(-q,-\Omega)\right)\\
&\qquad\qquad\qquad\quad+
G_\Lambda^A(\omega,k)G_\Lambda^R(\omega+\Omega,k+q)
\Tr_E\left(\widehat P_-a_t(q,\Omega)\widehat P_+a_t(-q,-\Omega)\right)
\Big]\\
&=-\frac{D_t}{4}\int_{q,\Omega}
\Tr_E\left([\Lambda,a_t(q,\Omega)][\Lambda,a_t(-q,-\Omega)]\right).
\end{aligned}
\end{equation}
Here, we have used the identity $\Tr(\widehat P_+a_1\widehat P_-a_2+\widehat P_-a_1\widehat P_+a_2)=-\frac14\Tr([\Lambda,a_1][\Lambda,a_2])$ and taken the slow-field limit $G_\Lambda^{R/A}(\omega+\Omega,k+q)\simeq G_\Lambda^{R/A}(\omega,k)$. The coefficient $D_t$ is
$$
D_t=\int_{k,\omega}G_\Lambda^R(\omega,k)G_\Lambda^A(\omega,k)
=\int_{-\pi}^{\pi}\frac{dk}{2\pi}\int_{-\infty}^{\infty}\frac{d\omega}{2\pi}
\frac{1}{(\omega-\xi_k)^2+(\gamma+w)^2}
=\frac{1}{2(\gamma+w)}.
$$
In real space, the second-order temporal-gradient contribution is $-\frac{1}{16(\gamma+w)}\int dt\,dx\,\Tr([\Lambda,a_t]^2)$.

For the spatial quadratic term, an analogous derivation, together with the slow-field approximations $v_{k+q}\simeq v_k$ and $G_\Lambda^{R/A}(\omega+\Omega,k+q)\simeq G_\Lambda^{R/A}(\omega,k)$, gives
\begin{equation}
\begin{aligned}
\Tr(G_\Lambda\hat v a_xG_\Lambda\hat v a_x)
&=-\frac{D_x}{4}\int_{q,\Omega}
\Tr_E\left([\Lambda,a_x(q,\Omega)][\Lambda,a_x(-q,-\Omega)]\right),\\
D_x&=\int_{k,\omega}v_k^2G_\Lambda^R(\omega,k)G_\Lambda^A(\omega,k)
=\frac{1}{2(\gamma+w)}\int_{-\pi}^{\pi}\frac{dk}{2\pi}v_k^2
=\frac{1}{\gamma+w}.
\end{aligned}
\end{equation}
Here, we have used $\overline{v^2}\equiv\int_{-\pi}^{\pi}\frac{dk}{2\pi}v_k^2=2$ for the one-dimensional nearest-neighbor hopping Hamiltonian. The second-order spatial-gradient contribution is therefore $-\frac{D_x}{8}\int dt\,dx\,\Tr([\Lambda,a_x]^2)$, and the two contributions combine into
\begin{equation}
iS^{(2)}[Q]=-\frac18\int dt\,dx\,\Tr_E\left(D_t[\Lambda,a_t]^2+D_x[\Lambda,a_x]^2\right).
\end{equation}

We define $A_\mu^+=U^{-1/2}\partial_\mu U^{1/2}$ and $A_\mu^-=U^{1/2}\partial_\mu U^{-1/2}$, so that $a_\mu=\hat P_+^xA_\mu^++\hat P_-^xA_\mu^-$. Using the previously fixed Keldysh representation of the saddle, we have $[\Lambda,a_\mu]=i\sigma_K^y(A_\mu^+-A_\mu^-)$, with $A_\mu^+-A_\mu^-=U^{-1/2}(\partial_\mu U)U^{-1/2}$. After carrying out the Keldysh trace, it follows that
$$
\Tr_E\left([\Lambda,a_\mu]^2\right)
=-2\Tr_R\left[U^{-1}(\partial_\mu U)U^{-1}(\partial_\mu U)\right]
=2\Tr_R\left[(\partial_\mu U^\dagger)(\partial_\mu U)\right].
$$
where we have traced over the $2\times 2$ Keldysh space, and $\Tr_R$ denotes the trace over the $R\times R$ replicated space. 
The trace-log contribution to the $SU(R)$ NLSM is therefore
\begin{equation}\label{eq:NLSM1}
iS^K[U]=-\frac{1}{4}\Tr\left[
D_t(\partial_tU^\dagger)(\partial_tU)
+D_x(\partial_xU^\dagger)(\partial_xU)
\right].
\end{equation}
where we redefine $\Tr$ as the trace over the internal $(x,t)$ space and the $R\times R$ replica space.
We next consider fluctuations away from the $t=t'$ saddle. Following the convention used for the $t=t'$ sector, we first parametrize the full field as $T=e^{i\sigma_K^x\Phi/2}=\hat{P}^x_+U^{1/2}+\hat{P}^x_-U^{-1/2}$, with $U=e^{i\Phi}$, so that $Q=T\Lambda T^{-1}=\hat{P}^x_+\Lambda\hat{P}^x_-U+\hat{P}^x_-\Lambda\hat{P}^x_+U^\dagger$. Applying the nonequal-time projector to this identity gives $Q^{t\neq t'}=\hat{P}^x_+\Lambda\hat{P}^x_-U^{t\neq t'}+\hat{P}^x_-\Lambda\hat{P}^x_+(U^\dagger)^{t\neq t'}$. For a projector symmetric under $t\to t'$, one has $(U^\dagger)^{t\neq t'}=(U^{t\neq t'})^\dagger$, and the Keldysh-projector algebra then gives $\Tr[(Q^{t\neq t'})^2]=2\Tr[U^{t\neq t'}(U^{t\neq t'})^\dagger]$. We note that the full field $U$ is unitary, $UU^\dagger=\mathbf 1$, whereas its $t=t'$ and $t\neq t'$ projections are not individually unitary.

Since the two projected temporal sectors are orthogonal under the full trace, $\Tr[Q^2]=\Tr[(Q^{t=t'})^2]+\Tr[(Q^{t\neq t'})^2]$. Then we combine the $t=t'$ contribution in Eq.~(\ref{eq:saddleAction3}) with the $t\neq t'$ contribution in Eq.~(\ref{eq:FreqAction}), and simplify with $\Tr\left[\left(Q_{(\pm)}^{t\neq t'}\sigma_K^z\right)^2\right]=
\Tr\left[\left(Q_K^{t\neq t'}\right)^2\right]$, where the lower $(\pm)$ indices refer to the standard Keldysh $\pm$ basis before Keldysh rotation. The common term proportional to $\Tr(Q^2)$ is constant, while the difference between the two coefficients yields the nontrivial contribution,
\begin{equation}
i\Delta S_K^{t\neq t'}[U]=-\frac{(\gamma+w)\gamma}{2 w}\Tr\left[U^{t\neq t'}(U^{t\neq t'})^\dagger\right].
\end{equation}

We now split the field into fast and slow components, $U=U_fU_0$, with $U_f=e^{i\Phi_f}$. The slow background $U_0$ is constructed around the saddle $Q_s=\Lambda$, which is stationary, carries external frequency $\Omega=0$, and is supported in the equal-time sector. Stationarity places the saddle contribution in the Wilsonian slow sector, while its equal-time structure implies that it is annihilated by the non-equal-time projector. Since the slow field is dominated by the saddle and its nearby configurations, we make the leading
saddle-point approximation $\mathcal P_{t\neq t'}U_0\simeq0$, where $\mathcal P_{t\neq t'}$ denotes the projector onto the non-equal-time sector. We let $\Phi_{f}^{t\neq t'}\equiv\mathcal P_{t\neq t'}\Phi_f$
denote the non-equal-time component of the Wilsonian fast field. Expanding the fast field to linear order, we obtain,
\begin{equation}
U^{t\neq t'}=\mathcal P_{t\neq t'}(U_fU_0)=\mathcal P_{t\neq t'}(U_0)
+i\mathcal P_{t\neq t'}(\Phi_fU_0)+O(\Phi_f^2)\simeq i\Phi_{f}^{t\neq t'}U_0+O(\Phi_f^2).
\end{equation}
We neglect the $O(\Phi_f^2)$ corrections because they do not
contribute to the quadratic fast-field propagator. The approximation
$\mathcal P_{t\neq t'}U_0\simeq0$ is an additional saddle-dominance
approximation accompanying the usual Wilsonian shell decomposition:
the retained slow field is approximated by its equal-time component,
whereas its non-equal-time deviations are not kept as explicit
running fields. Their effects are absorbed into local matching
corrections after the non-equal-time sector is integrated out.
Using the unitarity of $U_0$, the mass term becomes, to quadratic order in the fast field,
\begin{equation}
i\Delta S_K^{t\neq t'}[U]\simeq-\frac{(\gamma+w)\gamma}{2w}\Tr\left(
\Phi_f^{t\neq t'}U_0U_0^\dagger(\Phi_f^{t\neq t'})^\dagger\right)
=-\frac{(\gamma+w)\gamma}{2w}\Tr\left(\Phi_f^{t\neq t'}(\Phi_f^{t\neq t'})^\dagger\right).
\end{equation}
Thus, the non-equal-time part of the fast propagator acquires a finite nonderivative quadratic term. Below the corresponding mass scale, it has no infrared singularity and can be integrated out. Its integration may generate finite local renormalizations of the bare NLSM couplings, but it does not control the infrared RG flow of the remaining equal-time $t=t'$ sector. We therefore omit this massive sector from the subsequent RG analysis.

To cast the action in Eq.~(\ref{eq:NLSM1}) into a more uniform form, we use $[D_t]=T/L$ and $[D_x]=L/T$ and introduce the characteristic velocity $v_*=\sqrt{D_x/D_t}=\sqrt{2}$. The action can then be written as
\begin{equation*}
iS^K[U]=-\frac{\sqrt{2}}{8(\gamma+w)}\Tr\left[
\frac{1}{v_*}(\partial_tU^\dagger)(\partial_tU)
+v_*(\partial_xU^\dagger)(\partial_xU)
\right].
\end{equation*}
Defining the rescaled coordinates $\tilde{t}=v_*t$ and $\tilde{x}=x$, and transforming both the derivatives and the integration measure contained in $\Tr$, we obtain the standard isotropic NLSM form
\begin{equation}
iS^K[U]=-\frac{\sqrt{2}}{8(\gamma+w)}\Tr\left[
\sum_{\mu=\tilde{x},\tilde{t}}(\partial_\mu U^\dagger)(\partial_\mu U)
\right].
\end{equation}
As a result, after integrating out the massive sectors, the remaining theory is the $SU(R)$ NLSM, $iS^K[U]=-\frac{D}{2}\Tr\sum_\mu(\partial_\mu U^\dagger)(\partial_\mu U)$, with $D=\sqrt{2}/[4(\gamma+w)]$, where $D$ denotes the leading-order bare stiffness at the initial RG scale of the remaining slow-mode NLSM. Here, we first resolve the trace as $\Tr=\int d\tilde{x}\,d\tilde{t}\,\Tr_R$, then choose the embedded Hermitian generators $\hat T^a$ of the $SU(R)$ replicon sector, normalized by $\Tr_R(\hat T^a\hat T^b)=\delta^{ab}$, with $[\hat T^a,\hat T^b]=if^{abc}\hat T^c$ and $f^{acd}f^{bcd}=2R\delta^{ab}$.
Let $\xi^a$, $a=1,\ldots,R^2-1$, be local coordinates on $SU(R)$, with $U=U(\xi)$ and $\partial_a\equiv\partial/\partial\xi^a$. Using $\partial_\mu U=(\partial_aU)\partial_\mu\xi^a$, we define the target-space metric by $g_{ab}(\xi)\equiv\Tr_R[(\partial_aU^\dagger)(\partial_bU)]$, so that $\Tr_R[(\partial_\mu U^\dagger)(\partial_\mu U)]=g_{ab}(\xi)\partial_\mu\xi^a\partial_\mu\xi^b$. This metric is invariant under both left and right multiplication of $U$ and therefore defines the bi-invariant metric on $SU(R)$. Since $\mathfrak{su}(R)$ is simple, this invariant quadratic form is proportional to the Killing form; with the present normalization, $g_{ab}$ is the normalized positive Killing metric and satisfies $g_{ab}(0)=\Tr_R(\hat T^a\hat T^b)=\delta^{ab}$ at the identity.
Defining the positive Euclidean functional by $S_E\equiv-iS^K$, the matrix action takes the standard geometric form, 
\begin{equation}
S_E=\frac{1}{2}\int d\tilde{x}\,d\tilde{t}\,G_{ab}(\xi)\sum_{\mu=\tilde{x},\tilde{t}}\partial_\mu\xi^a\partial_\mu\xi^b \qquad G_{ab}(\xi)=Dg_{ab}(\xi).
\end{equation}
In the background-field method, we decompose the field as $U=e^{i\xi^a\hat T^a}U_0$, where $U_0$ is a slowly varying background and $\xi^a$ are chosen as Riemann normal coordinates centered at $U_0$. In these coordinates, the metric has the local expansion $g_{ab}(\xi)=g_{ab}(0)-\frac{1}{3}R_{acbd}\xi^c\xi^d+\cdots$. The Gaussian contraction of the fast fields contracts the Riemann tensor over the fast tangent-space indices, producing the Ricci-tensor correction to the target-space metric.
For the general geometric NLSM above, the one-loop Wilsonian flow towards the infrared is $\partial_s G_{ab}=-(2\pi)^{-1}\operatorname{Ric}_{ab}[G]$, where $s\equiv\log(\Lambda_{\text{RG},0}/\Lambda_\text{RG})$, and $\Lambda_{\text{RG},0}$ and $\Lambda_\text{RG}$ denote the initial and running cutoffs of external momentum, respectively \cite{Friedan1980,Friedan1985}. For the bi-invariant $SU(R)$ metric, the Ricci curvature is isotropic in the target-space tangent directions, with $\operatorname{Ric}_{ab}[g]=\frac{1}{4}f^{acd}f^{bcd}=\frac{R}{2}g_{ab}$. Comparing the geometric flow with $G_{ab}=Dg_{ab}$ therefore gives
\begin{equation}\label{eq:RG_SU_R}
\left. \frac{dD}{d\log l}\right|_{\mathrm{AIII}}=-\frac{R}{4\pi}+O(D^{-1}),
\qquad
D(l_0)\equiv D_0=\frac{\sqrt{2}}{4(\gamma+w)},
\end{equation}
where we have defined the RG length scale $l$ by $s=\log(l/l_0)$, $l_0=\Lambda_{\text{RG},0}^{-1},l=\Lambda_\text{RG}^{-1}$.
Since the Ricci tensor is proportional to the bi-invariant metric itself, all tangent directions remain renormalized isotropically, so the RG flow is completely described by the single running stiffness $D$.

For the clean monitored system, the additional PHS$^\dagger$ constraint places the replicon sector in class BDI, with the NLSM target manifold $SU(2R)/Sp(2R)$. The corresponding one-loop RG equation has been derived in Ref.~\cite{poboiko2025a}. In the present normalization, the BDI flow is given by,
\begin{equation}\label{eq:BDI}
\left.\frac{dD}{d\log l}\right|_{\mathrm{BDI}}
=-\frac{R}{2\pi}+O(D^{-1}).
\end{equation}
Thus, the BDI stiffness flows towards strong coupling twice as fast as the AIII stiffness, and the corresponding exponential correlation-length scale in AIII is twice as large in the exponent.

To begin with the analysis of correlation length, we identify the microscopic length scale at which the NLSM becomes applicable. The dressed Green functions have the form $G_\Lambda^{R/A}(\omega,k)=[\omega-\xi_k\pm i(\gamma+w)]^{-1}$, and the retarded--advanced kernel satisfies $\int\frac{d\omega}{2\pi}G_\Lambda^R(\omega,k)G_\Lambda^A(\omega,k)=1/[2(\gamma+w)]$. We therefore define the microscopic coarse-graining time $\tau_0=1/[2(\gamma+w)]$. The slow-field expansion requires $|\Omega|\tau_0\ll1$ and $|q|v_*\tau_0\ll1$,  and hence the initial RG length scale of the NLSM is 
$$
l_0\equiv v_*\tau_0=\frac{\sqrt{2}}{2(\gamma+w)}.
$$
Since the monitored system is heated to infinite temperature, $v_*$ is the root-mean-square velocity averaged over the full Brillouin zone. Accordingly, $l_0$ should be interpreted as the microscopic coarse-graining length set by the retarded--advanced broadening.

For finite disorder, we introduce the crossover length $l_w$, below which the coarse-grained theory does not yet resolve the symmetry-breaking effect of disorder. If $l_{\rm cor}<l_w$, the flow reaches strong coupling before the crossover and is governed entirely by the BDI equation. If $l_0<l_w<l_{\rm cor}$, the flow is initially BDI for $l_0<l<l_w$ and crosses over to AIII for $l_w<l<l_{\rm cor}$. For any finite $w$, the exact symmetry classification remains AIII; the "BDI flow" here denotes only the preasymptotic RG regime in which the explicit symmetry breaking has not yet been resolved. Finally, when $l_w\lesssim l_0$, disorder is already resolved at the microscopic matching scale and the entire flow is governed by AIII.

This two-stage flow can be made explicit by defining $D_0=D(l_0)$ and using the one-loop criterion $D(l_{\rm cor})\simeq0$. The pure-class correlation lengths at the same initial stiffness are $l_{\rm cor}^{\mathrm{BDI}}=l_0\exp(2\pi D_0/R)$ and $l_{\rm cor}^{\mathrm{AIII}}=l_0\exp(4\pi D_0/R)$. We then consider the crossover regime $l_w\in (l_0,l_{\rm cor})$ and perform the two-stage integration,
$$
 D_0-D(l_{\rm cor})= 2a \log\left(\frac{l_w}{l_0}\right) +a \log\left(\frac{l_{\rm cor}}{l_w}\right)\quad \rightarrow\quad 
 \log\left(\frac{l_{\rm cor}^\mathrm{AIII}}{l_0} \right) = \log\left(\frac{l_{\rm cor}}{l_0} \right)+\log\left( \frac{l_w}{l_0}\right),
$$ 
where $a=\frac{R}{4\pi}$. We can write down the expression for $l_{\rm cor}$ in different regimes,
\begin{equation}\label{eq:crossover_lcor}
l_{\rm cor}=
\begin{cases}
l_{\rm cor}^{\mathrm{BDI}},
& l_w\geq l_{\rm cor}^{\mathrm{BDI}},\\[2mm]
l_{\rm cor}^{\mathrm{AIII}}\dfrac{l_0}{l_w},
& l_0<l_w<l_{\rm cor}^{\mathrm{BDI}},\\[3mm]
l_{\rm cor}^{\mathrm{AIII}},
& l_w\leq l_0.
\end{cases}
\end{equation}
Here, since for $l_w\geq l_{\rm cor}$ the RG flow is governed entirely by the BDI equation in Eq.~(\ref{eq:BDI}), we replace the condition $l_w\geq l_{\rm cor}$ by the equivalent condition $l_w\geq l_{\rm cor}^{\mathrm{BDI}}$.
In the crossover regime, we find $l_{\rm cor}=l_{\rm cor}^{\mathrm{AIII}}\dfrac{l_0}{l_w}$, and the BDI segment therefore reduces the pure-AIII correlation length only by the multiplicative factor $l_0/l_w$. Physically, since the subsequent AIII flow accumulates twice as much logarithmic length for the same remaining stiffness, the leading exponential dependence is already governed by the AIII coefficient once the crossover occurs in the bulk of the crossover regime, even when the disorder is weak.

To justify the AIII exponential dependence in the cross-over regime, $l_w$ should not itself depend exponentially on $(\gamma+w)^{-1}$. To estimate the scaling of the cross-over length $l_w$, we regard finite disorder as a weak explicit symmetry-breaking perturbation of the clean BDI NLSM. Locally around the surviving AIII sector, the tangent space of the BDI manifold can be decomposed into directions tangent to the AIII manifold and transverse directions that are soft only in the clean BDI theory. We parametrize the $U$ field near the AIII manifold by $U=U_{\parallel}e^{X_\perp}$, where $U_{\parallel}\in SU(R)$ parametrizes the surviving AIII sector, while $X_{\perp}$ parametrizes the transverse directions that are present only in the BDI manifold. We expect the following leading-order correction for the transverse modes $X_\perp$, $S_\perp^{(2)}=\frac12\int d^2r\,\left[D(\partial_\mu X_\perp)^2+\kappa_w X_\perp^2\right]$. Here, finite disorder, treated as a weak perturbation to the clean system, induces the correction $S_w$, whose first-order variation with respect to $X_\perp$ vanishes at $X_\perp=0$, while the second-order transverse Hessian $\left.\frac{\partial^2 S_w}{\partial X^\perp_i\partial X^\perp_j}\right|_{X_\perp=0}$ gives rise to the mass term with strength $\kappa_w$. The mass term leads to a finite mass in the transverse propagator $G_\perp(q)\propto\frac{1}{Dq^2+\kappa_w}=\frac{1}{D}\frac{1}{q^2+l_w^{-2}}$, which defines the cross-over length $l_w\equiv\sqrt{\frac{D}{\kappa_w}}$. The coefficient $\kappa_w$ can generally depend on both $\gamma$ and $w$. As $\kappa_w$ is generated at tree level by the leading perturbative symmetry-breaking correction, no essential exponential dependence is generated at this order, so that $l_w$ does not modify the leading exponential factor in the asymptotic limit.

As the disorder strength $w$ is increased within the crossover regime, $l_w$ decreases and the initial BDI interval becomes progressively shorter. The correlation length consequently increases towards its nearly pure-AIII value. Once $l_w\lesssim l_0$, this symmetry-crossover enhancement has saturated because the flow is AIII from the matching scale onward. Further increasing disorder then mainly acts through the replacement $\gamma\to\gamma'=\gamma+w$, which decreases the bare stiffness $D_0\propto(\gamma+w)^{-1}$. The correlation length therefore reaches a maximum close to the pure-AIII regime and is subsequently suppressed as $w$ is increased further.

Finally, to connect the replicon NLSM with the physical density correlation, we follow the source construction of Ref.~\cite{poboiko2023}. The replica-traceless density source couples directly to the $SU(R)$ replicon field, and differentiating the generating functional with respect to this source yields the replica-off-diagonal equal-time density correlation. At the Gaussian level, one obtains $C(q)\simeq D(l_0)|q|$, which gives the algebraic decay $C(r)\propto-r^{-2}$ at noncoincident points. The different NLSM target manifolds do not affect this Gaussian result, so both AIII and BDI exhibit the same leading $r^{-2}$ decay. Their difference appears only beyond the Gaussian level through the class-dependent renormalization of the stiffness $D$. At scales $r\gtrsim l_{\rm cor}$, the perturbative NLSM enters the strong-coupling regime. Following Ref.~\cite{poboiko2023} and the analogy with Anderson localization, we expect the algebraic density correlation to cross over to an exponential decay, $C(r)\sim \exp(-r/l_{\rm cor})$.

\end{document}